\documentclass{aastex62}

\received{}
\revised{}
\accepted{July 15, 2019}

\shorttitle{Volatile Composition of C/2016 R2 (PanSTARRS)}
\shortauthors{McKay et al.}

\usepackage[modulo, pagewise]{lineno}
\usepackage{ulem}
\usepackage{color,soul}

\begin{document}

\title{The Peculiar Volatile Composition of CO-Dominated Comet C/2016 R2 (PanSTARRS)}

\correspondingauthor{Adam McKay}
\email{adam.mckay@nasa.gov}

\author{Adam J. McKay}
\altaffiliation{Visiting Astronomer at the Infrared Telescope Facility, which is operated by the University of Hawaii under contract NNH14CK55B with the National Aeronautics and Space Administration.}
\affil{NASA Goddard Space Flight Center\\
8800 Greenbelt Rd\\
Greenbelt, MD, 20771 USA}
\affiliation{Department of Physics\\
American University\\
4400 Massachusetts Avenue NW\\
Washington, D.C., 20016 USA}

\author{Michael A. DiSanti}
\altaffiliation{Visiting Astronomer at the Infrared Telescope Facility, which is operated by the University of Hawaii under contract NNH14CK55B with the National Aeronautics and Space Administration.}
\affiliation{NASA Goddard Space Flight Center\\
8800 Greenbelt Rd\\
Greenbelt, MD, 20771 USA}

\author{Michael S. P. Kelley}
\affiliation{Department of Astronomy\\
University of Maryland\\
4296 Stadium Dr.\\
College Park, MD, 20742 USA}

\author{Matthew M. Knight}
\affiliation{Department of Astronomy\\
University of Maryland\\
4296 Stadium Dr.\\
College Park, MD, 20742 USA}

\author{Maria Womack}
\affiliation{Florida Space Institute\\
University of Central Florida\\
Partnership 1, Research Parkway\\
Orlando, FL, 32816 USA}

\author{Kacper Wierzchos}
\affiliation{Department of Physics\\
University of South Florida\\
4202 East Fowler Ave\\
Tampa, FL, 33620 USA}

\author{Olga Harrington Pinto}
\affiliation{Department of Physics\\
University of South Florida\\
4202 East Fowler Ave\\
Tampa, FL, 33620 USA}

\author{Boncho Bonev}
\affiliation{Department of Physics\\
American University\\
4400 Massachusetts Avenue NW\\
Washington, D.C., 20016 USA}

\author{Geronimo L. Villanueva}
\affiliation{NASA Goddard Space Flight Center\\
8800 Greenbelt Rd\\
Greenbelt, MD, 20771 USA}

\author{Neil Dello Russo}
\altaffiliation{Visiting Astronomer at the Infrared Telescope Facility, which is operated by the University of Hawaii under contract NNH14CK55B with the National Aeronautics and Space Administration.}
\affiliation{Johns Hopkins University Applied Physics Laboratory\\
11100 Johns Hopkins Rd.\\
Laurel, MD, 20723 USA}

\author{Anita L. Cochran}
\affiliation{University of Texas at Austin/McDonald Observatory\\
2512 Speedway, Stop C1402\\
Austin, TX, 78712 USA}

\author{Nicolas Biver}
\affiliation{LESIA, Observatoire de Paris, Universit\'e PSL, CNRS,\\ 
Sorbonne Universit\'e, Universit\'e de Paris, Sorbonne Paris Cit\'e\\
5 place Jules Janssen, F-92195 Meudon, France\\}

\author{James Bauer}
\affiliation{Department of Astronomy\\
University of Maryland\\
4296 Stadium Dr.\\
College Park, MD, 20742 USA}

\author{Ronald J. Vervack, Jr.}
\affiliation{Johns Hopkins University Applied Physics Laboratory\\
11100 Johns Hopkins Rd.\\
Laurel, MD, 20723 USA}

\author{Erika Gibb}
\affiliation{Department of Physics\\
1 University Blvd.\\
University of Missouri-St.Louis\\
St. Louis, MO, 63121 USA}

\author{Nathan Roth}
\affiliation{Department of Physics\\
1 University Blvd.\\
University of Missouri-St.Louis\\
St. Louis, MO, 63121 USA}

\author{Hideyo Kawakita}
\affiliation{Koyama Astronomical Observatory\\
Kyoto Sangyo University\\
Motoyama, Kamigamo, Kita-ku, Kyoto, 603-8555, Japan}





\begin{abstract}
Comet C/2016 R2 (PanSTARRS) has a peculiar volatile composition, with CO being the dominant volatile as opposed to H$_2$O and one of the largest N$_2$/CO ratios ever observed in a comet.  Using observations obtained with the \textit{Spitzer Space Telescope}, NASA's Infrared Telescope Facility, the 3.5-meter ARC telescope at Apache Point Observatory, the Discovery Channel Telescope at Lowell Observatory, and the Arizona Radio Observatory 10-m Submillimeter Telescope we quantified the abundances of 12 different species in the coma of R2 PanSTARRS: CO, CO$_2$, H$_2$O, CH$_4$, C$_2$H$_6$, HCN, CH$_3$OH, H$_2$CO, OCS, C$_2$H$_2$, NH$_3$, and N$_2$.  We confirm the high abundances of CO and N$_2$ and heavy depletions of H$_2$O, HCN, CH$_3$OH, and H$_2$CO compared to CO reported by previous studies.  We provide the first measurements (or most sensitive measurements/constraints) on H$_2$O, CO$_2$, CH$_4$, C$_2$H$_6$, OCS, C$_2$H$_2$, and NH$_3$, all of which are depleted relative to CO by at least one to two orders of magnitude compared to values commonly observed in comets.  The observed species also show strong enhancements relative to H$_2$O, and even when compared to other species like CH$_4$ or CH$_3$OH most species show deviations from typical comets by at least a factor of two to three.  The only mixing ratios found to be close to typical are CH$_3$OH/CO$_2$ and CH$_3$OH/CH$_4$.  The CO$_2$/CO ratio is within a factor of two of those observed for C/1995 O1 (Hale-Bopp) and C/2006 W3 (Christensen) at similar heliocentric distance, though it is at least an order of magnitude lower than many other comets observed with AKARI.  While R2 PanSTARRS was located at a heliocentric distance of 2.8 AU at the time of our observations in January/February 2018, we argue, using sublimation models and comparison to other comets observed at similar heliocentric distance, that this alone cannot account for the peculiar observed composition of this comet and therefore must reflect its intrinsic composition.  We discuss possible implications for this clear outlier in compositional studies of comets obtained to date, and encourage future dynamical and chemical modeling in order to better understand what the composition of R2 PanSTARRS tells us about the early Solar System. 
\end{abstract}



\section{Introduction} \label{sec:intro}
Comets are primitive, volatile-rich remnants from the formation of the Solar System, and for this reason their volatile composition is considered indicative of the physics and chemistry occurring in the protosolar disk during the planet formation stage.  While there is diversity in compositions among the cometary population, comets consist primarily of H$_2$O, followed by CO$_2$ and CO at the 1-30\% level, with trace species such as HCN, CH$_3$OH, and C$_2$H$_6$ being present at the few percent level or less~\citep{MummaCharnley2011}.  The optical taxonomy of comets also shows diversity and evidence for modality (i.e. clumps of different compositional types), but in general optical cometary spectra are dominated by OH (and therefore H$_2$O), with CN, C$_2$, C$_3$, CH, NH, and NH$_2$ present at the percent level or less~\citep[e.g.][]{AHearn1995, Cochran2012}.\\
\indent However, a few comets do not fit neatly into the established taxonomies.  Comet 96P/Machholz shows an extremely atypical coma composition at optical wavelengths (though OH is still the dominant species), with a C$_2$/CN ratio an order of magnitude higher than other comets~\citep{LanglandShulaSmith2007, Schleicher2008}.  Some comets, such as 29P/Schwassman-Wachmann 1, have observed CO/H$_2$O ratios $>>$ 1~\citep{Ootsubo2012}, though this is often, at least partially, explained by its heliocentric distance being beyond the water ice line where water ice does not readily sublimate, meaning the observed coma composition is not necessarily indicative of the ice composition of the nucleus~\citep{WomackDistantReview2017}.\\
\indent Comet C/2016 R2 (PanSTARRS) was observed to have a peculiar optical spectrum when it was at a heliocentric distance of 3.1 AU, dominated by emissions from CO$^+$ and N$_2^+$ with a lack of emission from other species such as CN and C$_2$ typically observed at these wavelengths~\citep{CochranMcKay2018}.  From these observations, it was derived that N$_2$/CO $\sim$ 0.06, among the highest values observed in a comet.  This result, coupled with the lack of many of the usual emissions observed in optical cometary spectra, suggested that this comet has a composition very different from any other comet observed to date.  This was quickly communicated to the cometary science community for additional observations.  Observations with the Arizona Radio Observatory 10-m Submillimeter Telescope (ARO SMT) confirmed a very high CO production rate and identified a low HCN abundance~\citep{WierzchosWomack2017, WierzchosWomack2018, deValBorro2018}.  Observations with the IRAM 30-meter telescope and the Nancay radio telescope provided a more complete picture of the volatile composition of R2 PanSTARRS, confirming the high CO abundance and anomalously low abundances of other volatiles such as HCN and H$_2$O compared to CO~\citep{Biver2018}.  Additional high spectral resolution optical observations obtained with the UVES instrument on the VLT were reported by~\cite{Opitom2019}, confirming the findings of \cite{CochranMcKay2018} as well as providing new insights such as the first detection of [\ion{N}{1}] emission in a cometary coma.\\
\indent We present analysis of IR measurements obtained with the \textit{Spitzer Space Telescope} and iSHELL on the NASA IRTF designed to quantify a suite of species observed in comets: H$_2$O, CO$_2$, CO, C$_2$H$_6$, CH$_3$OH, CH$_4$, H$_2$CO, and OCS.  We also present optical measurements used to study HCN (through CN emission), NH$_3$ (through NH$_2$ emission), N$_2$ (through N$_2^+$), C$_2$H$_2$ (through C$_2$) and H$_2$O (through OH and [\ion{O}{1}]6300~\AA~emission), as well as new millimeter-wavelength observations of CO that are contemporaneous with our \textit{Spitzer} observations. Section 2 presents our observations  and Section 3 presents our analysis procedures and results.  Section 4 discusses this truly peculiar comet in the context of current compositional taxonomies and possible implications for the physics and chemistry of the comet-forming region during the protoplanetary disk phase.  Section 5 concludes the paper and encourages future work to better understand what R2 PanSTARRS reveals about the early Solar System.

\section{Observations} \label{sec:obsda}
\indent We obtained observations in late January/February 2018 with several facilities, both space-borne and ground-based.  These observations are detailed in Table~\ref{observations}.

\begin{table}[h!]
\begin{center}
\caption{\textbf{Observation Log}
\label{observations}
}
\begin{tabular}{cccccccc}
\hline
UT Date & R$_h$ (AU) & $\dot{R_h}$ (km s{$^{-1}$}) &$\Delta^a$ (AU) & $\dot{\Delta}^a$ (km s{$^{-1}$}) & Solar Standard & Tell. Standard & Flux Standard\\
\hline
NASA IRTF iSHELL & & & & & & &\\
January 30, 2018 & 2.81 & -6.8 & 2.27 & +17.2 & - & HR 1165 & HR 1165\\
\hline
ARO SMT & & & & & &\\
February 13, 2018 & 2.76 & -6.0 & 2.41 & +19.5 & - & - & Chopper Wheel\\
\hline
\textit{Spitzer} IRAC$^b$ & & & & & & &\\
February 12, 2018 & 2.76 & -6.0 & 2.26 & -28.4 & - & - & -\\
February 21, 2018 & 2.73 & -5.5 & 2.11 & -25.9 & - & - & -\\
\hline
DCT LMI & & & & & &\\
February 21, 2018 & 2.73 & -5.5 & 2.52 & +20.3 & - & - & HD 72526\\
 & & & & & & & HD 37112\\
\hline
APO ARCES & & & & & &\\
January 30, 2018 & 2.81 & -6.8 & 2.27 & +17.2 & Hyades 64 & 55 Persei & HR 1544
\end{tabular}
\end{center}
$^a$Values for $\Delta$ and $\dot{\Delta}$ are the distance to the observer and velocity relative to the observer, respectively.  Therefore for \textit{Spitzer} observations these are relative to the \textit{Spitzer} spacecraft, while for ground-based observations these are relative to Earth.\\
$^b$\textit{Spitzer} astronomical observation request (AOR) numbers: 65280768, 65281280, 65281024, 65281536
\end{table}

\subsection{NASA IRTF iSHELL}
\indent We obtained Director's Discretionary Time to observe R2 PanSTARRS with the powerful iSHELL IR spectrograph on the NASA IRTF on Maunakea, HI on UT January 29 and 30, 2018.  While poor weather precluded obtaining useful data on January 29, we obtained high quality spectra on January 30.  The iSHELL detector is a 2048 $\times$ 2048 pixel Hawaii H2RG array with sensitivity over a wavelength range of $\sim$ 1-5 $\mu$m.  As a cross-dispersed instrument, iSHELL measures signal in many ($>$10) consecutive echelle orders simultaneously with complete (for $\lambda \leq 4$ $\mu$m) or nearly complete (for $\lambda > 4$ $\mu$m) spectral coverage, a significant improvement over the previous IRTF high-resolution facility spectrograph, CSHELL~\citep{Tokunaga1990}.  More details on iSHELL can be found in~\cite{Rayner2012, Rayner2016}.\\ 
\indent For our observations of R2 PanSTARRS, we used the 0.75\arcsec~wide slit, which provides a spectral resolution of R $\equiv$ $\frac{\lambda}{\delta\lambda}$ $\sim$ 38,000 for a uniform monochromatic source.  We also observed an early type IR standard star with the 4\arcsec~wide slit to serve as a flux calibrator and a telluric standard (see Section~\ref{subsec:ishellda}).  This slit provides lower spectral resolution (R $\sim$ 20,000), but minimizes slit losses and therefore systematic errors in flux calibration.  Both the comet and standard star were observed using the classic ABBA nodding sequence, with a 7.5\arcsec~telescope nod (half the slit length) along the slit between the A and B positions, located equidistant to either side of the slit midpoint.  We employed two grating settings: M2 and Lp1.  M2 covers a wavelength range of $\sim$ 4.52 - 5.25 $\mu$m, encompassing spectral lines of CO, H$_2$O, and OCS.  Lp1 covers the wavelength range $\sim$ 3.28-3.65 $\mu$m and targets emission from CH$_4$, C$_2$H$_6$, CH$_3$OH, H$_2$CO, and OH prompt emission (a well established proxy for H$_2$O production in comets~\citep{Bonev2006}).  Our comet observations resulted in 29.2 minutes on source in M2 and 59.8 minutes on source in Lp1.  We obtained flats and darks at the end of each observing sequence for each grating setting.\\  
\indent Guiding was achieved through filter imaging with the slit-viewer camera, performed in specific wavelength bands independent of the wavelength regime used to obtain spectra. The slitviewer allows active guiding on sufficiently bright targets while obtaining spectra.  Short time-scale guiding is achieved through a boresight guiding technique, which utilizes ``spillover'' flux that falls outside the slit to keep the optocenter on the slit.  However, while easily visible in the guider using a broadband J filter, R2 PanSTARRS was not bright enough for active guiding.  We instead performed offset guiding using a reference (guide) star in the slitviewer field of view (FOV).  Owing to the small non-sidereal rates of R2 PanSTARRS, this worked very well; we verified that the comet's position with respect to the slit remained very stable over the course of our observations, with minimal adjustments needed to keep it in the slit.  More details relevant to cometary observations using iSHELL are presented in~\cite{DiSanti2017}.\\

\subsection{Arizona Radio Observatory Sub-millimeter Telescope (SMT)}\label{subsec:smt}
Observations of the CO(2-1) line at 230.53799 GHz were performed with the Arizona Radio Observatory 10-m Submillimeter Telescope on 2018 February 13 at 5:16 UT using the 1.3 mm dual polarization receiver with ALMA Band 6 sideband-separating mixers.  The observations began 11 hours after the first \textit{Spitzer} epoch.  Data acquisition was done in beam-switching mode with a +2\arcmin~throw in azimuth and standard 6 minute scans were acquired. System temperatures had peak values of 390K, but on average remained under 340K. The chopper wheel method was used to determine the temperature scale for the SMT receiver systems with a beam efficiency of $\eta$ = 0.74. The backend configuration that provided the best velocity resolution (0.325 km s$^{-1}$ per channel) consisted of a 2048 channel 250 kHz/channel filterbank in parallel mode. Accuracy of the pointing and tracking was checked against the JPL Horizon’s ephemeris position and was found to be better than $<$ 1\arcsec~rms. Due to high winds, only 12 scans were obtained but the CO line was strong enough to be seen in single scans.

\subsection{\textit{Spitzer} IRAC}\label{subsec:spitzerobs}
\indent We obtained Director's Discretionary Time to observe R2 PanSTARRS with \textit{Spitzer} IRAC on 2018 February 12 at 18:22 UT and 2018 February 21 at 01:03 UT in order to measure the CO$_2$ production rate.  As \textit{Spitzer} is well into its post-cryogenic mission, IRAC presently observes in two pass bands: one centered at 3.6 $\mu$m and the other at 4.5 $\mu$m.  Both filters have broad wavelength coverage, with bandwidths of 0.8 and 1.0 $\mu$m, respectively.  The 4.5 $\mu$m band has been used extensively in the past for measuring CO$_2$ production rates in comets, as this bandpass includes the $\nu_3$ band of CO$_2$ at 4.26 $\mu$m~\citep[e.g.][]{Reach2013, McKay2016}.  It also contains the $\nu$(1-0) band of CO at 4.7 $\mu$m, but in many comets in the AKARI survey~\citep{Ootsubo2012}, the CO$_2$ feature was at least 10 times brighter than the CO feature, and so CO$_2$ is typically assumed to be the dominant gas emission feature in the IRAC 4.5 $\mu$m band.  This is due to the fluorescence efficiency of CO$_2$ being approximately an order of magnitude larger than that for CO coupled with the fact that the CO$_2$ abundance in comets is often equal to or greater than the CO abundance. There are examples, however, such as C/2006 W3 (Christensen) and 29P/Schwassman-Wachmann 1, where CO emission contributes significantly to the 4.5 $\mu$m band flux~\citep{Ootsubo2012,Reach2013}. The high CO production rate found for R2 PanSTARRS~\citep{WierzchosWomack2018,deValBorro2018, Biver2018} means CO emission may not be negligible in the Spitzer imaging, and may in fact dominate signal in the 4.5 $\mu$m channel.  We discuss how we account for the CO contribution in the Spitzer imaging in Section~\ref{subsec:spitzerda}.\\
\indent We supply details of our observations in Table~\ref{observations}.  The IRAC array is a 256 x 256 pixel InSb array, covering a 5.2\arcmin~x 5.2\arcmin~FOV with a spatial scale of 1.2\arcsec/pixel, which for our observations corresponds to a projected FOV of $\sim$500,000 km and a spatial scale of $\sim$1900 km/pixel.  We employed a 9-position random dither pattern.  Each bandpass is observed with independent arrays, such that when one array is observing the comet, the other is on an adjacent field.  The sequence takes about 12 minutes to execute.  For each epoch we performed observations of the comet field several days after each cometary observation in order to image the field without the comet in it.  These images are termed ``shadow observations'' and provide a measurement of the background to be subtracted from the cometary images.  Our observations were obtained in high dynamic range (HDR) mode, which entailed obtaining exposures with both short (1.2 s) and long (30 s) exposure times in order to avoid saturation of the inner coma, while still keeping high signal-to-noise ratio (SNR) in the fainter outer coma.  Observing in HDR mode also helps protect against saturation due to bright field stars.  For these observations no pixels were saturated; therefore to optimize SNR we analyzed the longest exposure time images.

\subsection{Discovery Channel Telescope-Large Monolithic Imager, (LMI)}
\indent We observed R2 PanSTARRS with the Large Monolithic Imager (LMI) on the 4.3-m Discovery Channel Telescope (DCT) at Lowell Observatory from 2:51--4:13 UT on February 21, 2018. The observations were a Target of Opportunity (ToO) request that interrupted normal observations and began within 2 hours of the {\it Spitzer} observations in order to provide a near-simultaneous constraint on the water production of R2 PanSTARRS. DCT ToO requests are limited to 2~hr, which allowed us sufficient time to focus the instrument, observe high and low airmass standard stars (listed in Table~\ref{observations}), and obtain $\sim$83 minutes on R2 PanSTARRS. Conditions were photometric with seeing $\sim$1.2\arcsec. The comet was $\sim$33$^\circ$ from a 26\% illuminated moon, although no stray light attributable to the moon was evident in any of our images.  The comet's airmass ranged from 1.06 to 1.23.

\indent LMI has a 12.3\arcmin$\times$12.3\arcmin{} field of view and $6.1 \mathrm{K} {\times} 6.1 \mathrm{K}$ e2v CCD. On-chip $3{\times} 3$ binning resulted in a pixel scale of 0.36\arcsec/pixel. We obtained images using a standard broadband SDSS-$r'$ filter and narrowband ion (CO$^+$ central wavelength/bandpass width=4266~\AA/64~\AA), gas (OH 3090/62, CN 3870/62), and dust continuum (UC 3448/84, BC 4450/67) filters that are all part of the comet Hale-Bopp set~\citep{Farnham2000}. Single frames were acquired at the start and end of the sequence in the two filters with the highest signal-to-noise, $r'$ (30 s) and CO$^+$ (300 s). In between, sets of exposures were acquired in OH (3 exposures each of 600 s), UC (3$\times$300 s), and BC (2$\times$120 s), with the sets proceeding from shortest to longest wavelength in order to minimize the effects of atmospheric extinction as the airmass increased. A single CN exposure (180 s) was also acquired at the end of the sequence. The telescope followed the comet's ephemeris rate for all comet images.

\subsection{Apache Point Observatory-ARCES}
\indent We obtained Director's Discretionary Time to observe R2 PanSTARRS with the ARCES instrument on the Astrophysical Research Consortium (ARC) 3.5-meter telescope at Apache Point Observatory (APO) in Sunspot, NM on UT January 30, 2018, just hours before the iSHELL observations.  ARCES provides a spectral resolving power of R = 31,500 and a spectral range of 3500-10,000~\AA~with no interorder gaps.  More specifics for this instrument are discussed elsewhere~\citep{Wang2003}.\\
\indent Observational details are described in Table~\ref{observations}.  For all observations we centered the 1.6\arcsec$\times$3.2\arcsec~slit on the optocenter of the comet.  We obtained six spectra of 1800 s each over the course of the night.  These spectra were averaged after extraction and calibration to increase SNR.  We obtained an ephemeris generated from JPL Horizons for non-sidereal tracking of the optocenter.  For short time-scale guiding, the guiding software employs a boresight technique to keep the optocenter centered in the slit.  We observed a G2V star in order to remove the underlying solar continuum and Fraunhofer absorption lines, a fast rotating (vsin(i) $>$ 150 km s$^{-1}$) B star to account for telluric features, and spectra of a flux standard to establish absolute intensities of cometary emission lines.  The calibration stars used are given in Table~\ref{observations}.  We obtained spectra of a quartz lamp for flat fielding and acquired spectra of a ThAr lamp for wavelength calibration.\\

\section{Data Analysis and Results} \label{sec:result}
\subsection{IRTF-iSHELL}\label{subsec:ishellda}
\begin{figure}[h!]
\begin{center}
\includegraphics[width=0.5\textwidth]{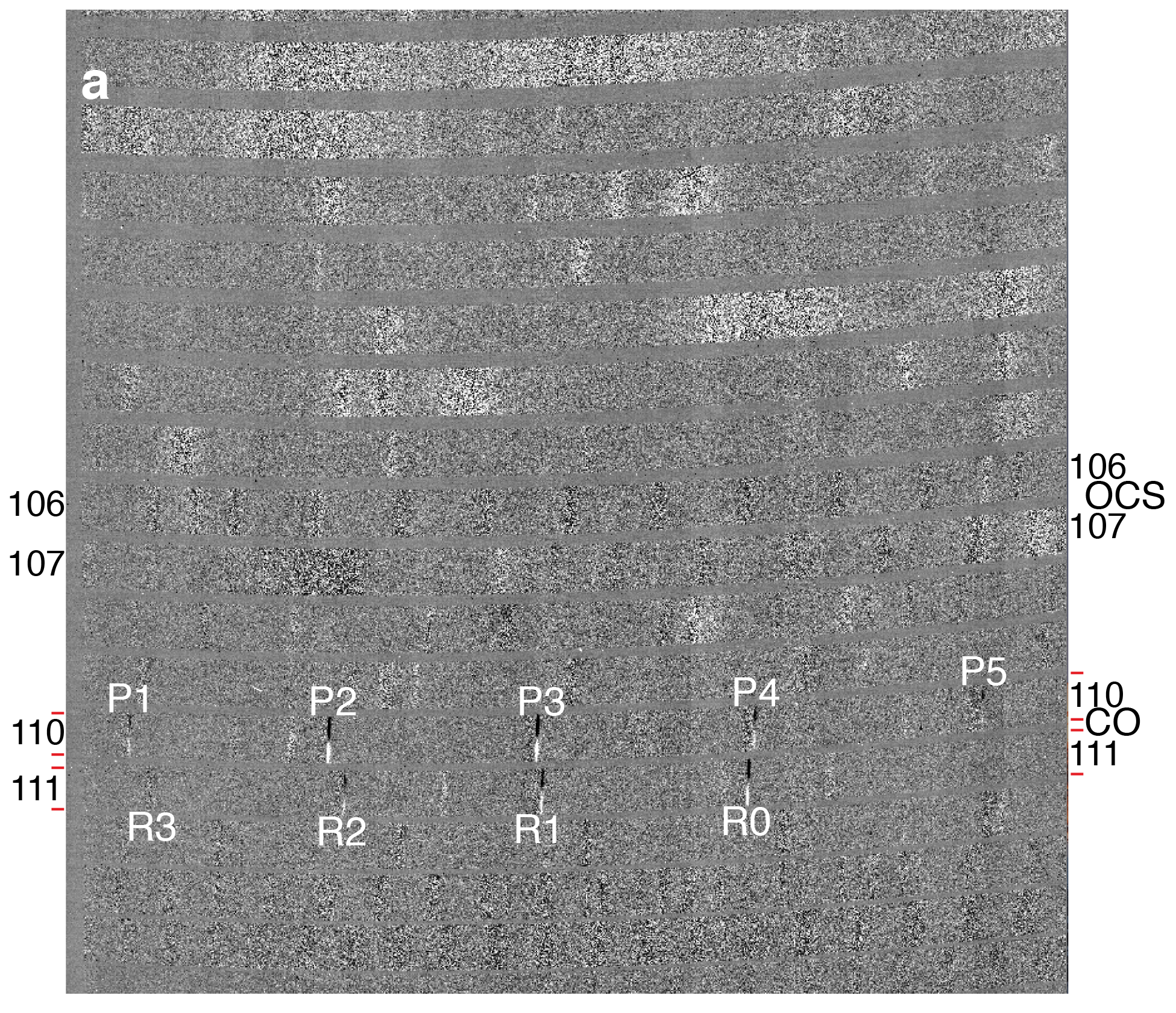}
\includegraphics[width=0.49\textwidth]{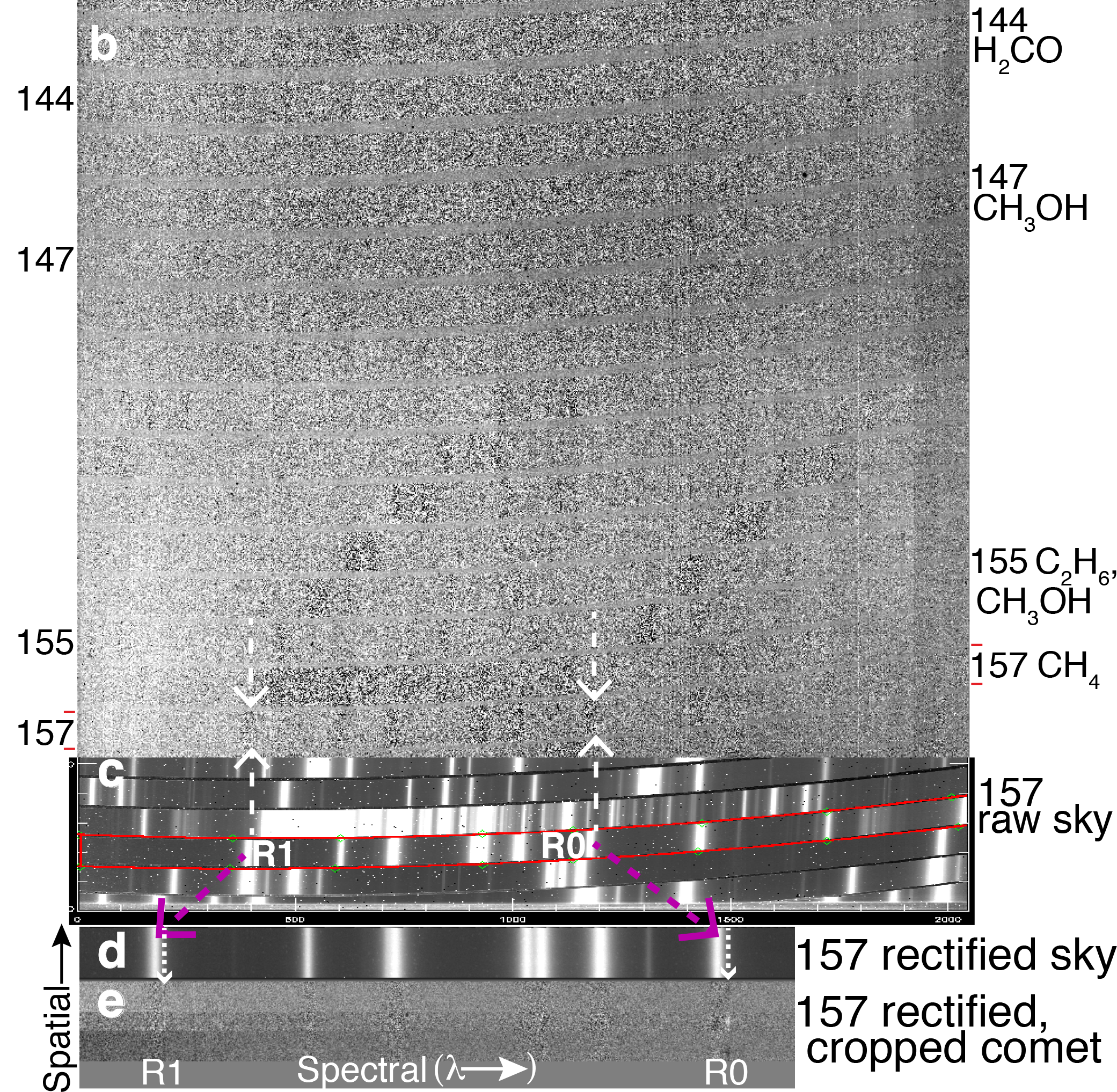}
\end{center}
\caption{
\label{iSHELL_stack}
Raw spectral-spatial stacks (A-B difference frames) of R2 PanSTARRS obtained with iSHELL in the (a) M2 and (b) Lp1 settings on UT 30 January 2018, with echelle orders relevant to our study indicated (106, 107, 110 and 111 for M2; 144, 147, 155, and 157 for Lp1), and with molecules targeted in each order listed at right.  CO emission lines (labeled with their rotational designations) are clearly seen in M2 orders 110 and 111, with the A-beam signal in white and the B-beam signal in black.  Conversely, no emission lines are readily apparent in the Lp1 stack.  (c) Portion of the corresponding Lp1 raw sky frame that includes order 157, outlined in red and spectrally aligned with the raw comet stack.  Expected positions of the cometary CH$_4$ lines are depicted by vertical white dashed arrows immediately to the right/red of the corresponding sky emission lines, in accordance with the geocentric velocity of R2 PanSTARRS ($\dot{\Delta} = +17.2$ km s$^{-1}$; Table 1). (d) Sky emission centered around the CH$_4$ lines in order 157, after being rectified as explained in the text.  (e) Corresponding rectified order for R2 PanSTARRS, again with the cometary CH$_4$ lines Doppler-shifted from their telluric counterparts. The combined comet signal (in white) is shown as the middle portion and, above and below this (in black) individual B- and A-beams, respectively.  For spectral extracts from the CO and CH$_4$ orders, see Fig.~\ref{iSHELL}.} 
\end{figure}

\indent Figure~\ref{iSHELL_stack} shows raw spectral-spatial difference frames (total A-beam minus B-beam exposures) of R2 PanSTARRS for our M2 (left) and Lp1 (upper right) observing sequences.  We applied our general methodology for processing IR spectra~\citep[e.g.][]{DelloRusso2006,Villanueva2011, DiSanti2014}.  New techniques specific to iSHELL are described in detail in~\cite{DiSanti2017} (see also~\cite{Roth2018}).  We provide a brief summary of our reduction procedures below.\\
\indent Processing of each iSHELL order produces a ``rectified'' spectral-spatial frame, meaning each column pertains to a unique wavelength and each row to a unique spatial location along the slit.  An example is shown in Figure~\ref{iSHELL_stack} for the region of Lp1 order 157 containing the cometary R0 and R1 lines of CH$_4$ (panel (e), bottom-right).  Such rectified orders consist of three parts: the bottom and top thirds show comet signal obtained from A-beam and B-beam observations, respectively (in black), and the middle third shows the combined signal [(A+B)/2] (in white).  Spectral extracts for the standard star and comet are obtained by summing signal over a range of rows in the central (combined-beam) portion of their rectified frames.\\

\begin{figure}[h!]
\includegraphics[width=\textwidth]{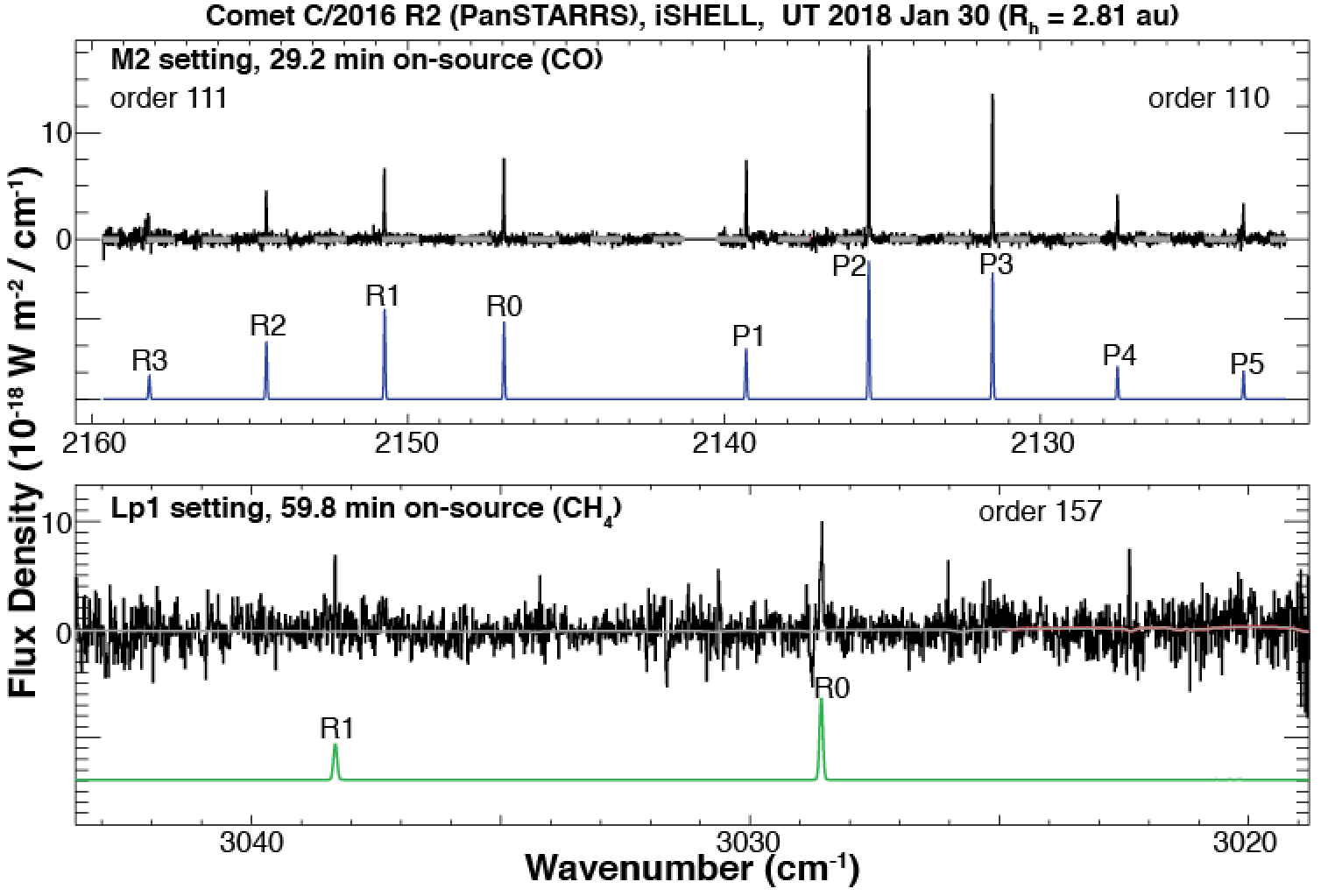}
\caption{
\label{iSHELL}
Spectra and corresponding fluorescence model fits to observed CO (top) and CH$_4$ (bottom) emission in R2 PanSTARRS obtained with iSHELL.  Each spectrum represents the signal summed over 15 rows (2.25'') centered on the peak emission, and each line is labeled with its rotational identification.  We detected nine CO lines spanning two echelle orders, and two CH$_4$ lines within a single order.}
\end{figure}

\indent To achieve absolute flux calibration, and to determine the column burdens of absorbing species in the terrestrial atmosphere, we fit a synthetic atmospheric transmittance model to the standard star spectrum for each processed order.  We applied this optimized atmospheric transmittance model (calculated at the airmass of R2 PanSTARRS), convolved it to the spectral resolution of the cometary observations and scaled it to the cometary continuum level (the continuum is virtually absent for our observations of R2 PanSTARRS).  Subtracting the scaled model yields the net observed cometary emission spectrum, still multiplied by monochromatic atmospheric transmittance at the Doppler-shifted frequency of each cometary line.  Correcting for transmittance and incorporating flux calibration factors from our standard star spectra allows establishing line fluxes incident at the top of the terrestrial atmosphere.   Fully calibrated spectral extracts for R2 PanSTARRS showing emission lines of CO and CH$_4$ are shown in Figure~\ref{iSHELL}.  All other species searched for were not detected.

\begin{figure}[h!]
\includegraphics[width=\textwidth]{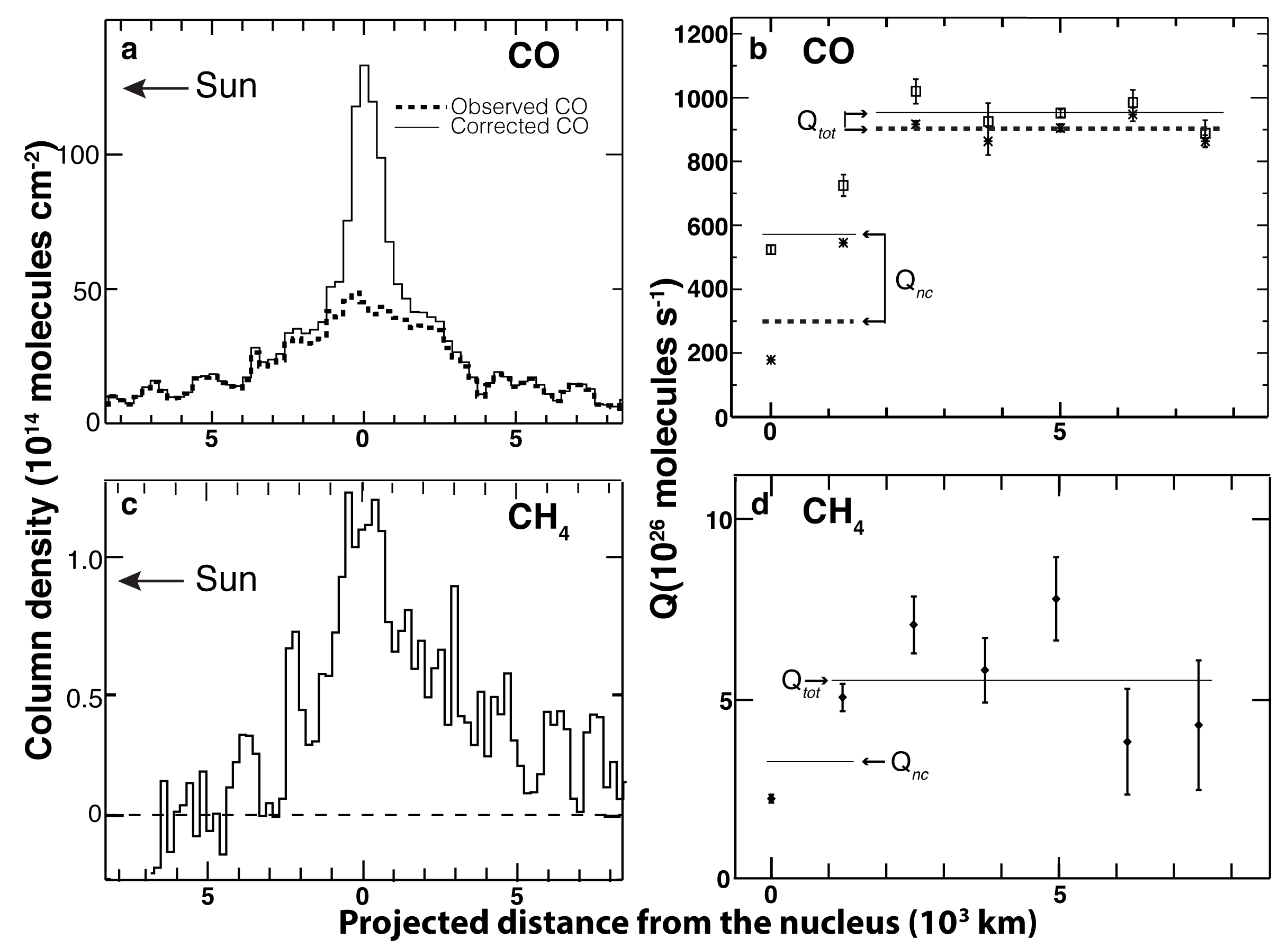}
\caption{
\label{Qcurve}
Spatial profiles (left) and symmetrized Q-curves (right)  from our iSHELL observations of R2 PanSTARRS, representing the sum of 9 CO lines (top) and the sum of 2 CH$_4$ lines (bottom).  For the spatial profiles, the sunward-facing hemisphere is to the left as indicated.  For CO (panels a and b), dashed traces/lines represent observed emission while solid traces/lines represent emission after correcting line-by-line for optical depth in the solar pump, following the methodology of~\cite{DiSanti2001}.  In all cases, GF is given by the ratio of total Q (Q$_{tot}$, upper horizontal line) to nucleus-centered Q (Q$_{nc}$, corresponding lower horizontal line).  The GF for CO prior to correcting for optical depth (2.93$\pm$0.04) is much larger than that measured for (optically thin) CH$_4$ (1.70$\pm$0.19).  However, the opacity-corrected CO Q-curve has a much lower GF = 1.67$\pm$0.04, consistent with that for CH$_4$.  We note that for CO our formalism for correcting optical depth increases Q$_{nc}$ by 91\%, but increases Q$_{tot}$ by only 5.6\% (well within 1-$\sigma$ uncertainty; see Table~\ref{Qrates}).  This suggests that our approach for modeling optical depth, coupled with the Q-curve formalism, provides robust production rates for these two species in R2 PanSTARRS. 
}
\end{figure}

\indent We establish molecular column densities (or upper limits) by dividing these transmittance-corrected line fluxes by appropriate line-specific fluorescence g-factors, the values of which depend on rotational temperature ($T_{rot}$). For our study of R2 PanSTARRS with iSHELL, only CO presented enough lines with high signal-to-noise that spanned a sufficient range of rotational energy to obtain a measure of $T_{rot}$. Because the solar spectrum contains CO absorption lines, an accurate treatment required using ``reduced'' CO g-factors that incorporate the Swings effect for the heliocentric velocity of the comet at the time of our observations (-6.8 km s$^{-1}$; see Table 1). Our analysis of CO provided a best-fit value $T_{rot}$ = 13$\pm$2 K.  We assume this temperature also applies to other species, as observations of brighter comets in which $T_{rot}$ was measured for multiple species demonstrate that this is generally a valid assumption~\citep[e.g.][]{DelloRusso2011, Mumma2011, Gibb2012, DiSanti2014}.\\
\indent We obtained molecular production rates as follows. We extracted a ``nucleus-centered'' spectrum by summing signal over 15 rows ($\sim$ 2.5\arcsec) centered on the row containing the peak emission line intensity. Application of Swings-corrected g-factors and geometric parameters (R$_h$, $\Delta$, beam size at the comet) to transmittance-corrected line fluxes provides the nucleus-centered production rate, Q$_{nc}$. However, owing primarily to seeing, Q$_{nc}$ invariably underestimates the actual ``total'' (or ``global'') production rate, Q$_{tot}$.  To obtain Q$_{tot}$ we multiplied each Q$_{nc}$ by an appropriate growth factor (GF), determined through the well-documented ``Q-curve'' method for analyzing spatial profiles of emissions~\citep{DelloRusso1998}.  For each spatial step, a ``symmetrized'' Q-curve was produced by averaging signal at equal but diametrically opposed distances from the nucleus.  For our observations, only CO and CH$_4$ were detected.  Each showed bright enough emissions to allow a reliable Q-curve analysis to be performed.\\  
\indent We present spatial profiles and symmetrized Q-curves for CO and CH$_4$ in Fig.~\ref{Qcurve}. For each Q-curve, GF is depicted graphically as the level of the upper horizontal line (representing Q$_{tot}$) divided by that of the corresponding lower horizontal line (representing Q$_{nc}$).  We refer to the region of the coma over which Q$_{tot}$ is measured as the ``terminal region.''

\indent Comparison of the profiles in Fig. 3 reveals a relatively broad (and in particular a ``flat-topped'') spatial distribution for the observed CO emission.  This is demonstrated by its much larger GF compared to CH$_4$, and also that leveling its Q-curve required beginning two steps from the nucleus instead of one step, as was used for CH$_4$.  This could indicate extended release of CO (e.g., from grains) in the inner coma, and/or optical depth in the CO lines (particularly along lines-of-sight passing close to the nucleus). The high CO production rates reported from millimeter observations of R2 PanSTARRS with IRAM \citep{Biver2018} and SMT \citep{WierzchosWomack2018}, as well as from our iSHELL observations (see Table 3) suggest the IR lines of CO are affected by optical depth.\\
\indent Several previous CO-rich comets revealed optically thick emissions that in all cases were most pronounced for lines-of-sight passing through the innermost coma. For observations of C/1995 O1 (Hale-Bopp) and C/1996 B2 (Hyakutake) with CSHELL at the IRTF, observed column densities and Q-curves were corrected for opacity in the solar pump assuming uniform gas outflow at constant speed~\citep{DiSanti2001, DiSanti2003}. Observations of C/2006 W3 (Christensen) with CRIRES at the ESO/VLT~\citep{Bonev2017} at a similar observing geometry to our observations of R2 PanSTARRS also revealed  optically thick CO for a production rate only slightly lower than has been reported for R2 PanSTARRS.~~\cite{Bonev2017} developed a formalism for addressing optical depth effects in CO emission based on a curve-of-growth analysis, demonstrating through their Q-curve analysis that the effects of optical depth on retrieved production rates can be quantified (and thus corrected for).\\
\indent Provided that signal in the terminal region (i.e., the ends of the slit farthest from the comet optocenter) approximates optically thin conditions, the Q-curve will level out, as was observed for these three previously observed CO-rich comets, and as we also observed for R2 PanSTARRS (Fig.~\ref{Qcurve}).  Although in this case for CO the central region is optically thick, the GF still allows establishing a reliable approximation of the actual Q$_{tot}$, but with the GF for CO being significantly larger than the corresponding value for the optically thin CH$_4$ emission in R2 PanSTARRS (see Fig. 3). For all these comets (including R2 PanSTARRS), Q$_{tot}$(CO) as retrieved from optically thin and optically thick treatments are in formal agreement (i.e., they agree to within their respective 1-$\sigma$ uncertainties); see Fig.~A4 of~\cite{DiSanti2001}, Fig.~4 of~\cite{DiSanti2003}, and Fig.~4 of~\cite{Bonev2017}. We demonstrate this by applying the methodology detailed in the Appendix of~\cite{DiSanti2001} to the observed CO emission in R2 PanSTARRS.  The presence of optically thick CO emission in R2 PanSTARRS is supported by the much larger GF for the observed CO profile (2.93 $\pm$ 0.04) versus that for (optically thin) CH$_4$ (1.70 $\pm$ 0.19), a discrepancy that is resolved by correcting for optical depth in the CO lines (resulting in 1.67 $\pm$ 0.04, see Fig.~3b).  Therefore we conclude that our Q-curve analysis for CO mitigates the effects of optical depth on our measured CO production rate.  It also reinforces our decision to apply its ``corrected'' GF (1.7) to obtain a realistic upper limit for co-measured OCS, and similarly to constrain Q$_{tot}$ using the (identical within 1-$\sigma$ uncertainty) GF from CH$_4$ for co-measured, undetected species included in the Lp1 setting (C$_2$H$_6$, CH$_3$OH, and H$_2$CO).  Derived production rates and mixing ratios are shown in Table~\ref{Qrates}.\\
\indent Our spatial profiles also permit testing for asymmetric gas outflow in the coma.  This is very pronounced for CH$_4$ (Fig. 3c), indicating a large enhancement in the anti-sunward facing hemisphere, particularly considering the relatively small solar phase angle of R2 PanSTARRS at the time of our observations ($\sim$ 19$^\circ$) and thus the potential high degree of projection onto the sky plane. The column density of CH$_4$ averaged between $\sim$ 2,000 and 8,000 km from the nucleus (projected on the sky) is a factor of 2.65$\pm$0.55 larger in the anti-sunward direction compared to the solar direction, while the CO profile is much more symmetric, its corresponding ratio being only 1.03$\pm$0.02.\\
\indent It is possible that the asymmetry observed for CH$_4$ is associated with rotation of the nucleus.  However, for the measured gas speed v$_{exp}$ of 0.52 km s$^{-1}$ (from our SMT observations, see Section~\ref{subsec:SMT-analysis}) the time required to exit the iSHELL terminal region (extending to 4.9'' from the nucleus) is approximately 4.3 hours, longer than the clock times encompassed by our M2 and Lp1 sequences (obtained consecutively and together spanning 2.65 hours of elapsed clock time).  This suggests that gas in the iSHELL slit was not replenished (at least significantly) between M2 and Lp1 observations, and thus the large differences in the degrees of asymmetry observed for CO and CH$_4$ cannot be explained by temporally variable outgassing.  Nonetheless, these results indicate that the abundance ratio CH$_4$/CO was higher in the anti-sunward-facing hemisphere than in the sunward-facing hemisphere by a factor of $\sim$ 2.5 and therefore, assuming little to no replenishment of coma gas between M2 and Lp1 sequences, this implies a significantly higher CH$_4$/CO abundance ratio in the anti-sunward direction at the time of our iSHELL observations.

\subsection{ARO-SMT}\label{subsec:SMT-analysis}
\begin{figure}[h!]
\includegraphics[width=\textwidth]{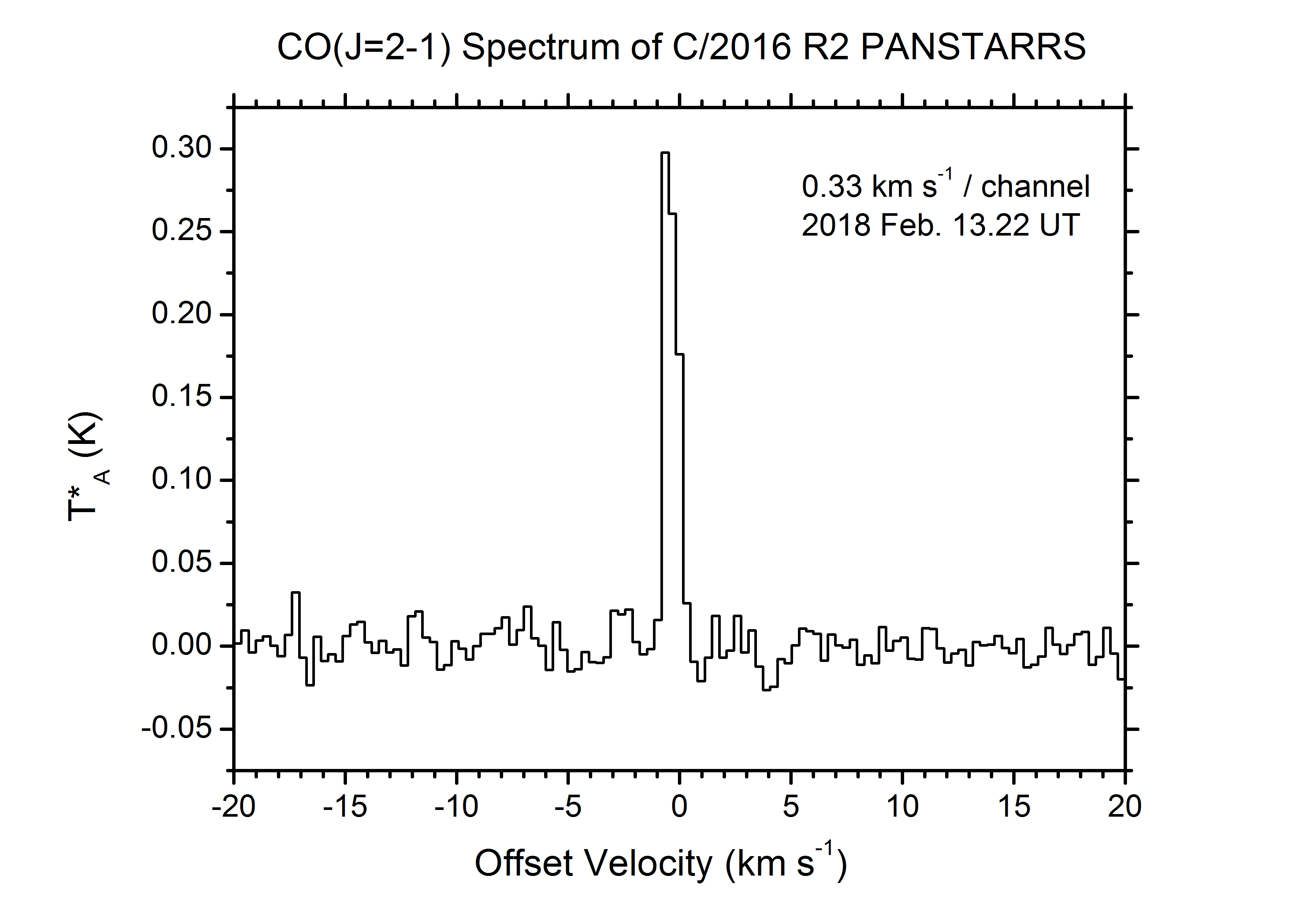}
\caption{
\label{SMT}
Spectrum of CO J=2-1 emission from comet C/2016 R2 obtained with the ARO SMT 10-m telescope on Feb. 13.22 2018 UT.}
\end{figure}

Figure~\ref{SMT} shows the spectrum resulting from coadding all 12 scans on UT February 13.  We calculated the column density assuming optically thin gas (appropriate for the relatively large beam diameter of 32$\arcsec$) with T$_{rot}$=23K~\citep[][see Section~\ref{sec:discuss} for discussion of difference with iSHELL]{Biver2018}, and we calculated the production rate assuming a simple symmetric outflow expansion model with V$_{exp}$=0.52 km s$^{-1}$ consistent with our spectra and other detailed modeling of the spectral line profiles \citep{WierzchosWomack2018,Biver2018}. We adopt this expansion velocity for all analysis in this work. The CO production rate is (5.5 $\pm$ 0.89) $\times$ 10$^{28}$ mol/s (see Table~\ref{Qrates}).
\subsection{\textit{Spitzer} IRAC}\label{subsec:spitzerda}
\begin{figure}[h!]
\begin{center}
\includegraphics[width=\textwidth]{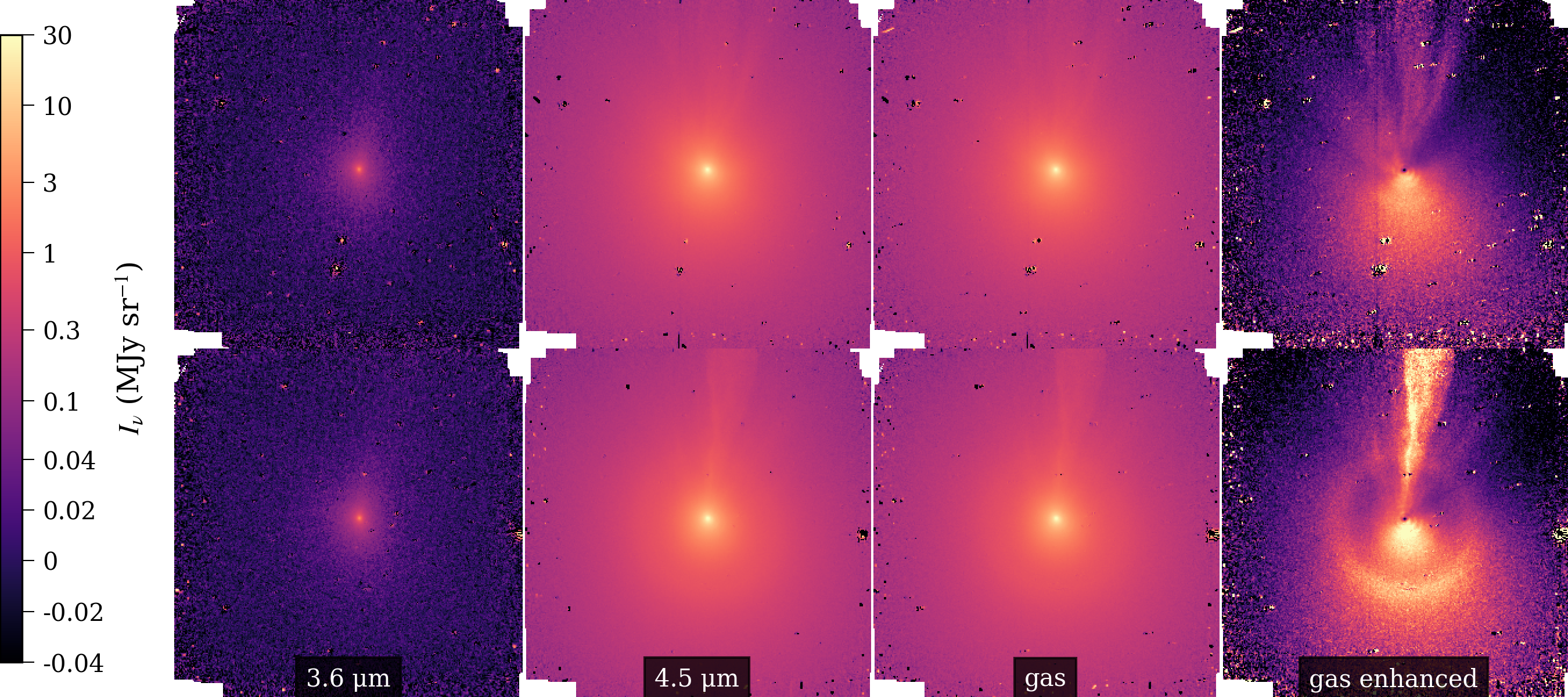}
\end{center}
\caption{
\label{Spitzer}
Shadow-subtracted images of C/2016 R2 (PanSTARRS) obtained with Spitzer IRAC at 3.6 $\mu$m (leftmost column) and 4.5 $\mu$m (left middle column), as well as the dust-subtracted images (right middle column) on UT February 12, 2018 (top row) and UT February 21, 2018 (bottom row).  These images are for the longer exposure time (30 s) in our HDR mode observations.  The solar direction is downward in all images.  The spherical coma geometry observed in the 4.5 $\mu$m channel, as well as the large flux ratio between the two filters, is indicative of a large gas contribution to the observed flux, due to CO$_2$ and CO.  The last column shows the dust-subtracted images normalized by a $\rho^{-1}$ surface brightness distribution, where $\rho$ is the projected distance to the peak brightness, which enhances coma asymmetries.  There is clear evidence for a strong ion tail, which we attribute to CO$^+$ (see Discussion). There is also a clear spiral structure in the inner-most coma indicative of a rotating jet and a large circular arc that is present on February 21 but not February 12, possibly indicative of an impulsive outgassing event in between the two observation epochs. On both dates there is broad sunward enhancement indicative of preferential outgassing from the sunward hemisphere, which is consistent with results from CO spectral line profiles and spatial mapping measured at millimeter wavelengths~\citep{WierzchosWomack2018, Biver2018}. A more detailed analysis of the coma morphology is beyond the scope of this paper and will be deferred to a future publication.}
\end{figure}

\indent We combined all images of the same exposure time using the MOPEX software \citep{Makovoz2005}.  This process creates a mosaic in the rest frame of the comet from the individual images, averaging overlapping data together, but ignoring cosmic rays and bad pixels.  Two mosaics are created: one for the comet data, the other for the shadow (background) data.  We subtracted the shadow mosaic from the comet mosaic to remove the background.  This includes zodiacal light and celestial sources.  While this removes background stars and most of the sky background, it may not completely remove the sky background (for instance, zodiacal light for a given RA and Declination varies with time).  We used the sky value from the adjacent image of blank sky to remove any residual background that remained after shadow subtraction.  The resulting images are shown in the first two columns of Fig.~\ref{Spitzer}.\\
\indent After the mosaic images were created and the sky background was subtracted, we removed the dust contribution from the 4.5 $\mu$m band flux, isolating the gas emission.  As the level of dust contamination is minimal as inferred from the 3.6 $\mu$m image being much fainter than the 4.5 $\mu$m image (see Fig.~\ref{Spitzer}), we simply applied a scaling factor of 0.9 to the 3.6 $\mu$m image and subtracted it from the 4.5 $\mu$m image to remove the dust contribution from the 4.5 $\mu$m image.  The scale factor of 0.9 was derived from a model for cometary dust accounting for the expected contributions of reflected light and thermal emission at the heliocentric distance of R2 PanSTARRS, based on the empirical coma dust model of \cite{Kelley2016} and the dust color of comet 67P/Churyumov-Gerasimenko at 2.8~au \citep{Snodgrass2017}.  The resulting dust-subtracted images are shown in the right center column of Fig.~\ref{Spitzer}.  The right-most column of Fig.~\ref{Spitzer} shows the dust-subtracted images normalized by a $\rho^{-1}$ surface brightness distribution, where $\rho$ is the projected distance to the peak brightness, which enhances coma asymmetries.  The enhanced images show strong spiral structures as well as an ion tail that we attribute to CO$^+$ (see Section~\ref{sec:discuss}).  A more detailed analysis of the coma morphology is beyond the scope of this paper.\\
\indent  As mentioned in Section~\ref{subsec:spitzerobs}, CO likely contributes significant flux to the \textit{Spitzer} 4.5 $\mu$m channel, especially considering the large CO production rates measured for R2 PanSTARRS at millimeter-wavelengths~\citep{WierzchosWomack2018,Biver2018} and in the IR with iSHELL (Fig. 3b).  CO$_2$ is also observed with strong emission at this wavelength in many comets \citep{Ootsubo2012}, with few other likely contributors (see Section~\ref{sec:discuss} for a discussion of other possible contaminating species). Therefore, we assume that the gas emission at 4.5 $\mu$m predominantly arises from CO and CO$_2$, and separate their contributions with a three-step process: 1) we initially assume that 100\% of the gas emission flux comes from CO molecules and derive a CO production rate, 2) subtract the contemporaneous ground-based measured CO production rate taking care to match the projected photometric aperture used for the Spitzer observations, which leaves a residual amount, and 3) then re-characterize this residual as a CO$_2$ production rate.\\
\indent From the dust-subtracted image of gas emission, we measured the flux for apertures ranging from 10-100 pixels (12-120\arcsec) in radius. We converted the broadband photometry to CO line fluxes in photons following the IRAC data handbook~\citep{Laine2015}.  The line fluxes were then used to calculate the total number of CO molecules inside the photometric aperture ($N_{CO,100\%}$) using
\begin{equation}
N_{CO,100\%}=4\pi\Delta^2F\frac{R_h^2}{g_{CO}}
\end{equation}
where $\Delta$ is the \textit{Spitzer}-comet distance, $F$ is the observed photon flux, $R_h$ is the heliocentric distance of the comet, and $g_{CO}$ is the g-factor for excitation of the CO 1$\rightarrow$0 fundamental vibrational band, which is 2.5 $\times$ 10$^{-4}$ photons s$^{-1}$~\citep{Debout2016}.  Then the production rate $Q_{CO,100\%}$, is given by
\begin{equation}
Q_{CO,100\%}=\frac{2v}{\pi \rho}N_{CO, 100\%}
\end{equation}
where $N_{CO,100\%}$ is the total number of CO molecules in the photometric aperture, $v$ is the expansion velocity, and $\rho$ is the projected radius of the photometric aperture.  We assume an expansion velocity of the coma of 0.52 km/s consistent with CO line widths from our SMT observations (see Section~\ref{subsec:SMT-analysis}) and with expansion velocities reported for R2 PanSTARRS by~\cite{WierzchosWomack2018} and~\cite{Biver2018}.  This approach assumes a negligible effect of photodissociation on the spatial profile in the photometric aperture, which is justified as our photometric apertures are $<$ 10\% of both the CO$_2$ and CO photodissociation scale lengths.  We calculated production rates for a variety of aperture sizes to quantify any trends in derived production rates with aperture size.  Since we found the deviations between derived production rates for different aperture sizes to be minimal ($<$ 5\%), for the rest of this paper we quote production rates derived using a photometric aperture that matches the projected distance at the comet of the ARO SMT observation beam (see Section~\ref{subsec:smt}). This is a 14-pixel radius on UT February 12 and a 15-pixel radius on UT February 21. This choice minimizes systematic errors when subtracting the expected CO flux from the \textit{Spitzer} images (see below).\\
\indent If we assume that all the gas flux in the \textit{Spitzer} 4.5 $\mu$m image is due to CO and calculate its corresponding production rate using equations (1) and (2), we find Q$_{CO, 100\%}$ $\sim$ 1.6 $\times$ 10$^{29}$ mol/s, higher than values derived from ground-based CO observations~\citep[][and this work]{WierzchosWomack2018, Biver2018, deValBorro2018}.  Therefore we conclude that CO is probably not the sole contributor to the observed flux. However, it is a major contributor and accounting for it does have a strong influence on the derived CO$_2$ production rate.  While CO contributes significantly to the observed \textit{Spitzer} fluxes, the excess we attribute to CO$_2$ is still $>$ 50\% of the observed flux, so the detection of CO$_2$ is robust.\\
\indent To calculate $Q_{CO_2}$ and account for the CO contribution to the \textit{Spitzer} 4.5 $\mu$m images, we employ our contemporaneous observations of CO at mm-wavelengths obtained with the ARO SMT (section~\ref{subsec:smt}).  We favor using the contemporaneous SMT results over the IRTF CO measurements from several weeks earlier for subtraction of the CO contribution from the \textit{Spitzer} imaging in order to minimize effects due to possible comet variability.  Moreover, the production rates from SMT and \textit{Spitzer} data were obtained using the same sized photometric aperture, whereas the iSHELL observations cover a much smaller projected region of the coma, making it more sensitive to short-time scale changes in gas production such as rotational variation.  This approach minimizes systematic uncertainties introduced by possible variability in the comet's activity.\\  
\indent We subtracted the mm-wavelength derived CO production rate, $Q_{CO,mm}$ (see Section~\ref{subsec:SMT-analysis}), from the derived $Q_{CO,100\%}$ value, which leaves a residual production rate: $Q_{residual} = Q_{CO,100\%}-Q_{CO,mm}$.  We attribute the residual gas production to CO$_2$ (see Section~\ref{sec:discuss} for a discussion of other possible contaminating species). This residual production rate is then converted to $Q_{CO_2}$ by taking advantage of the scaling relationship between production rates and fluorescence efficiencies: 
\begin{equation}
Q_{CO_2}=Q_{residual}{ {g_{CO}} \over {g_{CO_2}} }
\end{equation}
where $g_{CO_2}$ = 2.69 $\times$ 10$^{-3}$ photons s$^{-1}$ \citep{Debout2016}.  Using the SMT CO production rate, we derive Q$_{CO_2}$=(1.0 $\pm$ 0.1) $\times$ 10$^{28}$ mol/s (Table~\ref{Qrates}). As a demonstration of the sensitivity of our derived CO$_2$ production rate to the assumed value for CO, if Q$_{CO}$=0 then Q$_{CO_2}$ $\sim$ 1.5 $\times$ 10$^{28}$ mol/s, while Q$_{CO}$=1.0 $\times$ 10$^{29}$ mol/s (closer to the iSHELL value) results in Q$_{CO_2}$ $\sim$ 6.0 $\times$ 10$^{27}$ mol/s.\\
\indent Lastly, we derived the Af$\rho$ value as a proxy for the dust production based on the 3.6 $\mu$m flux levels using the same photometric aperture as for the gas photometry and assuming all the flux is solar continuum reflected off of dust particles in the coma (i.e., negligible emissions from gaseous species and no thermal emission).  We follow the methodology of~\cite{AHearn1984} to calculate Af$\rho$.  We derive Af$\rho$ values of 896 $\pm$ 27 cm on February 12 and 884 $\pm$ 27 cm on February 21, very low for a comet of this activity level and heliocentric distance.  We derive $log$[$Af\rho$/Q(H$_2$O)] = -23.54 $\pm$ 0.03, $log$[$Af\rho$/Q(CO)]= -25.79 $\pm$ 0.04, and $log$[$Af\rho$/Q(CO$_2$)]=-25.05 $\pm$ 0.04.  For similar reasons discussed above, we compare Af$\rho$ to the CO production rate measured by SMT rather than IRTF.\\
\indent Due to the high quality of our \textit{Spitzer} data, uncertainties in the photometry are dominated by the absolute calibration uncertainty of \textit{Spitzer} IRAC, which is approximately 3\%~\citep{Reach2005}.

\subsection{DCT-LMI}\label{subsec:DCT}
\begin{figure}[h!]
\begin{center}
\includegraphics[width=\textwidth]{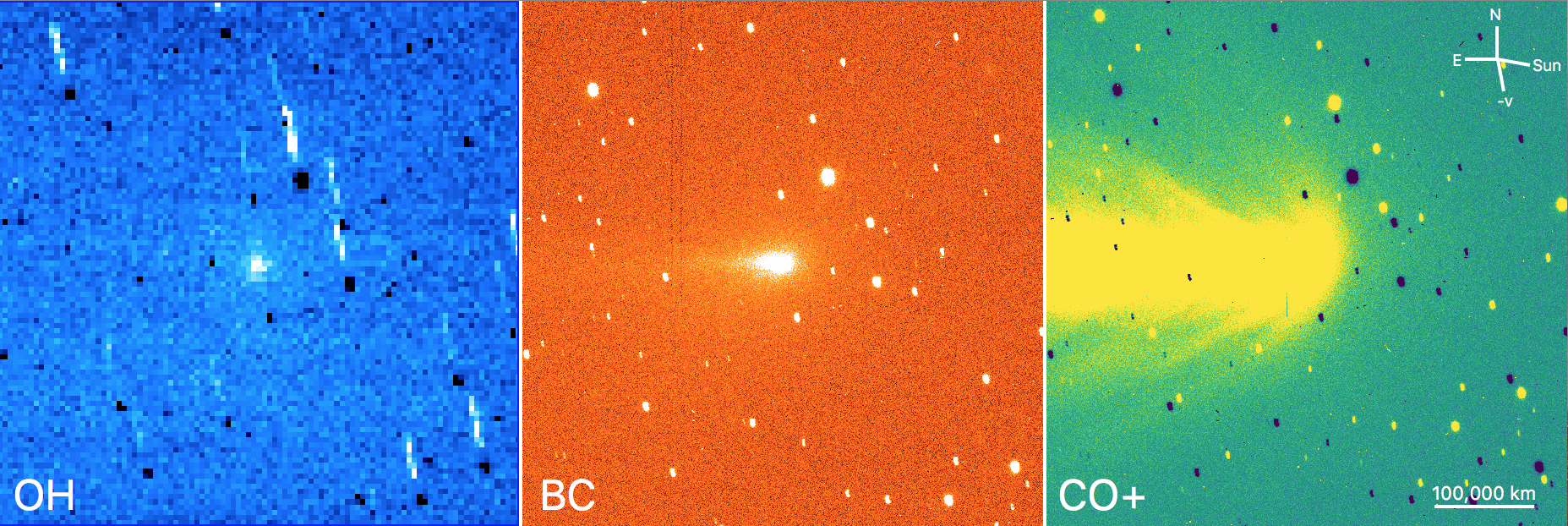}
\caption{
\label{fig:dct_images}
DCT images of OH (left, after dust and ion removal), BC (middle), and CO$^+$ (right, after dust removal). The OH image has been rebinned 8$\times$8 to improve signal to noise. Background stars are visible as trailed streaks in all images, while the negative of background stars (from the continuum image) are also visible in the OH and CO$^+$ images as black trailed streaks. Artifacts from a bad column on the CCD are faintly seen as vertical lines on the BC and CO$^+$ panels. A scale bar and arrows indicating north, east, and the directions to the Sun and the heliocentric velocity vector are given on the CO$^+$ panel.}
\end{center}
\end{figure}

The images were bias subtracted and flat-field corrected following standard practices, and all images for a given set (OH, UC, or BC) were median combined to improve signal-to-noise and mitigate background (stellar) contamination. Absolute calibrations and gas/dust decontamination for the narrowband filters were performed using extinction coefficients determined from the two standard stars and following the procedures outlined in \citet{Farnham2000}, with sky values determined from regions near the corners of the CCD that appeared to be visually free of gas/dust/ions. The CN and UC images exhibited extended, filamentary structures similar to those seen with the CO$^+$ filter. Although the filters were designed to isolate gas (CN) or to be a nearly emission-free continuum area (UC), the CN bandpass includes N$_2^+$ emission \citep{CochranMcKay2018,Opitom2019} and both bandpasses contain emissions from CO$^+$ ions \citep[][and references therein]{PearseGaydon1976}. For most comets, these ions are much fainter than the intended gas and/or continuum and can be safely ignored but this is not the case for R2 PanSTARRS. We thus conclude that the features we observed are ions and that UC and CN cannot be interpreted in their normal fashion. As a result, absolute calibrations of the OH images were performed using only the BC filter to define continuum, with solar color assumed.  Flux calibrated and decontaminated images are shown in Figure~\ref{fig:dct_images}.

\begin{figure}[h!]
\begin{center}
\includegraphics[width=0.3\textwidth]{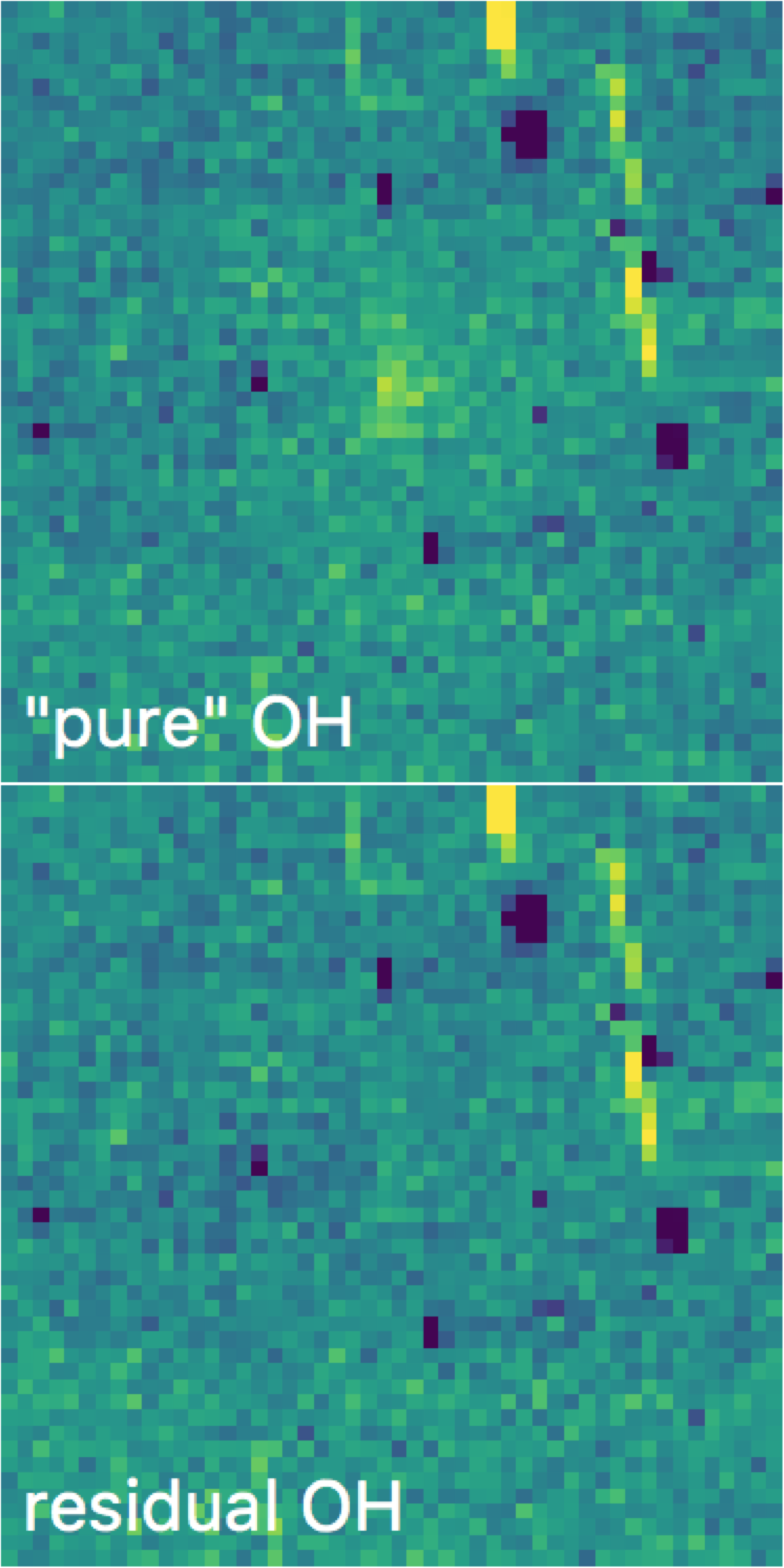}
\includegraphics[width=0.6\textwidth]{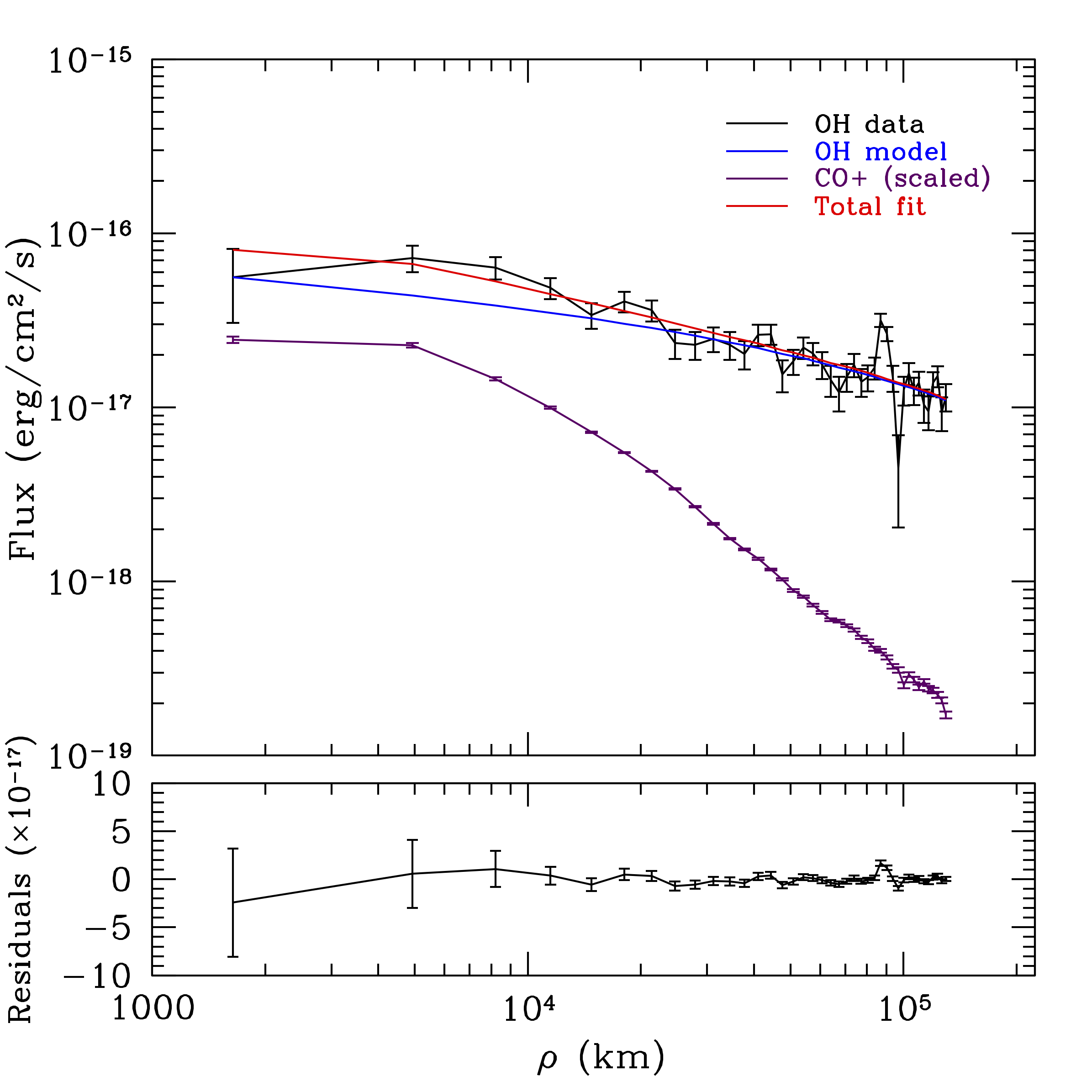}
\caption{
\label{fig:model_oh}
``Pure'' OH image produced using the scaling from our best-fit model (top left), residual OH after removing the modeled OH component from the ``pure'' OH image (top right), and the radial profiles in our best-fit model (bottom). The top panels are zoomed in from the OH image shown in Figure~\ref{fig:dct_images}, but have the same 8$\times$8 binning applied. They are each $\sim$260,000 km across at the comet, extending just wider than we fit the radial profiles shown in the bottom panel.
}
\end{center}
\end{figure}

Our normal procedure for determining gas production rates from fully calibrated images (e.g., \citealt{KnightSchleicher2015}) is to measure the flux in a circular aperture centered on the central condensation and convert to a production rate using a Haser model and standard assumptions about the appropriate lifetimes and scale lengths \citep{AHearn1995}. However, we elected to perform a more complex analysis, detailed below, due to several factors. First, the low signal-to-noise of our ``pure'' OH image meant that trailed stars or their negative residuals from the continuum image which was removed could significantly impact our inferred OH signal. Second, the aforementioned difficulties in decontaminating the image may have potentially led to over- or under-removal of sky background and/or underlying dust continuum. Finally, we identified four lines of CO$^+$ from the A$^{2}\Pi-\mathrm{X}^{2}\Sigma$ ``comet-tail system'' \citep[][and references therein]{PearseGaydon1976} that fall in the OH filter bandpass. While no ion tail structure is evident by eye in our OH image, we were nonetheless concerned about possible ion contamination that was not accounted for in the standard comet filter reduction procedures.\\ 
\indent In order to understand the extent to which our ``pure'' OH image might be contaminated by improperly removed continuum and/or ions, we extracted radial profiles from the uncalibrated CO$^+$, BC, and OH images after removing the background. The radial profiles were binned in distance from the nucleus, $\rho$, and in azimuthal angle, with a resistant mean used to screen out anomalously high or low pixel values. The dust and ion tails were oriented nearly due east, at a position angle (PA) of $\sim$90$^\circ$, slightly offset from the anti-sunward direction of PA = 79$^\circ$, so we determined radial profiles for the sunward and tailward hemispheres as well as 90$^\circ$ wedges centered at 0$^\circ$, 90$^\circ$, 180$^\circ$, and 270$^\circ$. We then fit slopes to the profiles from ${\rho}=4000-60000$~km (the inner radius was set to be about twice the seeing disk).  The exact slopes retrieved are very sensitive to the background level, but the behavior of slopes with azimuth is not.  The OH slope was the flattest, falling off as roughly ${\rho}^{-0.7}$ in all four quadrants. The BC slope was steeper, falling as roughly ${\rho}^{-1.1}$ in all four quadrants. The CO$^+$ exhibited the steepest slope and was the only one that varied significantly with PA, falling as ${\rho}^{-1.4}$ in the sunward quadrant, and ${\rho}^{-1.0}$ in the other quadrants. The CO$^+$ slopes are consistent with the expected behavior of ions whose sunward extent is minimal, while the consistency of the OH and BC slopes at all PAs suggest that each is relatively free of ion contamination.  Furthermore, the more rapid fall off of BC and CO$^+$ compared to OH suggests that, even if there were problems with improper decontamination of the OH image, the flux should increasingly approach the true OH signal at larger distances, with the sunward direction providing the least contaminated OH signal.\\
\indent This exercise also demonstrated that despite our previous concern about the low signal-to-noise of individual pixels in the OH image, with appropriate binning, clear signal could be detected to at least ${\rho}=1.2{\times}10^{5}$~km. Thus, we measured a radial profile in the sunward quadrant for the ``pure'' OH image and used this to determine the H$_2$O production rate on the assumption that the coma was spherically symmetric as follows. We created a synthetic OH profile using J.~Parker and M.~Festou's online version\footnote{{\tt http://www.boulder.swri.edu/wvm-2011/}} of the vectorial model \citep{Festou1981} using standard parameters given in Table~\ref{parameters} and scaling for the geometry of R2 PanSTARRS during our observations. We then interpolated the model to the midpoints of our radial profile $\rho$ bins and scaled it up or down to minimize ${\chi}^{2}$ from ${\rho}=4000-1.2{\times}10^{5}$~km. We repeated the process but added a second parameter, a fixed background offset that was allowed to be positive or negative, to minimize ${\chi}^{2}$ again. We continued to add complexity by including the BC and CO$^+$ profiles, first individually and then together, ultimately testing all combinations of vectorial model, fixed background, BC profile, and CO$^+$ profile. We repeated the fitting using an OH image that was flux calibrated with the continuum component set to zero as a further test of the reduction process.\\
\indent For the DCT imaging, the best-fit using all parameters is a water production rate, $Q$(H$_2$O), of $\sim$3$\times$10$^{26}$ mol/s. The best-fit for the BC scale factor was close to the solar color, implying 
that the dust was properly removed in the standard processing despite our earlier concerns.  Similarly, the best-fits for both images required a CO$^+$ component, supporting our suspicion that there was some ion contamination. In all cases, the fits yielded $Q$(H$_2$O) between 1 and 5$\times$10$^{26}$ mol/s, with the largest variations occurring for fits using only background fluxes (e.g., no continuum or ion component).  We tested reasonable deviations from the standard vectorial model assumptions, such as varying lifetimes and velocities, but these only changed $Q$(H$_2$O) at the $\sim$20\% level or less. The fits consistently show that the OH signal is much stronger than the CO$^+$ and dust continuum beyond $\sim$20,000 km.  Figure~\ref{fig:model_oh} shows a ``pure'' OH image using our best-fit parameters (top left panel), the residual flux after removing the modeled OH component from the ``pure'' OH image (top right), and the radial profiles in our best-fit model (bottom).

The residual OH image has been highly stretched to emphasize subtle features, but shows a hint of over-removal in the tailward direction and slight under-removal of the inner core.  Despite the large number of uncertainties and assumptions needed to deal with this unusual comet, we are confident that we have constrained $Q$(H$_2$O) to less than 10$^{27}$ mol/s with a most likely value of $\sim$(3.1$\pm$0.2)$\times$10$^{26}$ mol/s (Table~\ref{Qrates}).  The uncertainty quoted is strictly for our modeled assumptions and does not include systematic uncertainties that are difficult to quantify but could be larger.\\
\indent Using the observed flux in the BC image we calculated Af$\rho$ as described in Section~\ref{subsec:spitzerda}, although for the DCT data we used a smaller photometric aperture projected at the comet of 10,000 km, in line with the typical apertures employed by other optical dust photometry observations.  We obtain Af$\rho$=561 $\pm$ 6 cm (Table~\ref{Qrates}).\\
\subsection{APO-ARCES}\label{subsec:APO}
\indent Spectra were extracted and calibrated using IRAF scripts that perform bias subtraction, cosmic ray removal, flat fielding, and wavelength calibration.  We removed telluric absorption features, the reflected solar continuum from the dust coma, and flux calibrated the spectra employing our standard star observations.  We assumed an exponential extinction law and extinction coefficients for APO when flux calibrating the cometary spectra~\citep{Hogg2001}.  More details of our reduction procedures can be found in~\cite{McKay2012} and~\cite{CochranCochran2002}. We determined slit losses for the flux standard star observations by performing aperture photometry on the slit viewer images as described in~\cite{McKay2014}.  Slit losses introduce a systematic error in the flux calibration of $\sim$ 10\%.\\
\indent In our optical spectra we report analysis of five molecular species (N$_2^+$, CO$^+$, CN, NH$_2$, C$_2$) and one atomic species (OI).  Our analysis of CN, C$_2$, and NH$_2$ employs the same empirical fitting model employed by~\cite{McKay2014}, which utilizes a molecular line list from~\cite{CochranCochran2002} to fit Gaussian profiles to observed emission features.  For N$_2^+$ and CO$^+$ we adapt the model from~\cite{McKay2014} to include CO$^+$ by adding line positions from~\cite{Kuo1986,Haridass1992,Haridass2000} and N$_2^+$ line positions from~\cite{Dick1978}.  We integrate over the fits to measure the observed flux.  While we detect multiple bands of CO$^+$, we use the (2,0) band for analysis as it has the highest SNR.\\
\indent For CN, C$_2$, and NH$_2$ these fluxes are converted to production rates using a Haser model in which the input scale lengths are modified to emulate the vectorial model and g-factors from the literature.  Molecular lifetimes and g-factors employed are given in Table~\ref{parameters}.  For [\ion{O}{1}], we employ the observed [\ion{O}{1}]6300~\AA~emission as a proxy for H$_2$O production using a similar Haser model (see~\cite{McKay2012, McKay2014} for more details about the Haser models employed).  While [\ion{O}{1}]6300~\AA~emission is often used as a proxy for H$_2$O production in comets,~\citep[e.g.][]{Morgenthaler2001,Fink2009,McKay2018}, it assumes that H$_2$O is the dominant oxygen-bearing species in the coma, which is not the case for R2 PanSTARRS (see Table~\ref{Qrates}).\\
\indent Ions do not follow a Haser profile.  Therefore we do not calculate production rates from derived fluxes for N$_2^+$ and CO$^+$, only the relative ratio of N$_2^+$/CO$^+$ using

\begin{equation}
\frac{N_{N_2^+}}{N_{CO^+}}=\frac{F_{N_2^+}}{F_{CO^+}}\frac{g_{CO^+}}{g_{N_2^+}}
\end{equation}

\noindent where $N_x$ denotes column density of species $x$, $F_x$ is the observed flux of species $x$, and $g_x$ is the g-factor for the observed transition of species $x$.  The g-factors employed are $g_{CO^+}$=3.55 $\times$ 10$^{-3}$ photons s$^{-1}$ mol$^{-1}$~\citep{MagnaniAHearn1986} and $g_{N_2^+}$=7.00 $\times$ 10$^{-2}$ photons s$^{-1}$ mol$^{-1}$~\citep{Lutz1993}.

\begin{table}
\begin{center}
\caption{
\label{files}
\label{parameters}
}
\textbf{Parameter Values for Optical Species}\\

\begin{tabular}{lcccc}
Molecule & Parent Lifetime (s)$^a$ & Daughter Lifetime (s)$^a$ & g-factor (ergs s$^{-1}$ molecule$^{-1}$)$^a$\\
\hline
CN & 1.3 $\times$ 10$^4$ & 2.1 $\times$ 10$^5$& 2.6 $\times$ 10$^{-13}$\\
C2 & 2.2 $\times$ 10$^4$ & 6.6 $\times$ 10$^4$ & 4.5 $\times$ 10$^{-13}$\\
NH$_2$ & 4.1 $\times$ 10$^3$ & 6.2 $\times$ 10$^4$ & 6.40 $\times$ 10$^{-15}$\\
\ion{O}{1}$^b$ & 8.3 $\times$ 10$^4$ & - & -\\
\ion{O}{1}$^c$ & 1.3 $\times$ 10$^5$ & - & -\\ 
OH & 8.3 $\times$ 10$^4$ & 1.3 $\times$ 10$^5$ & 1.54 $\times$ 10$^{-15}$\\
\hline
\end{tabular}
\end{center}
$^a$ Given for $r$=1 AU.  For CN and OH, given for $\dot{r}$=0 km s$^{-1}$, but varies with $\dot{r}$.\\
$^b$ For [\ion{O}{1}] from dissociation of H$_2$O into H$_2$ and O; branching ratio employed is 0.07~\citep{BhardwajRaghuram2012}\\
$^c$ For [\ion{O}{1}] from dissociation of OH; branching ratio for H$_2$O to OH + H employed is 0.855~\citep{Huebner1992} and the branching ratio for OH to O + H is 0.094~\citep{BhardwajRaghuram2012}.\\
\end{table}

\begin{figure}[h!]
\includegraphics[width=\textwidth]{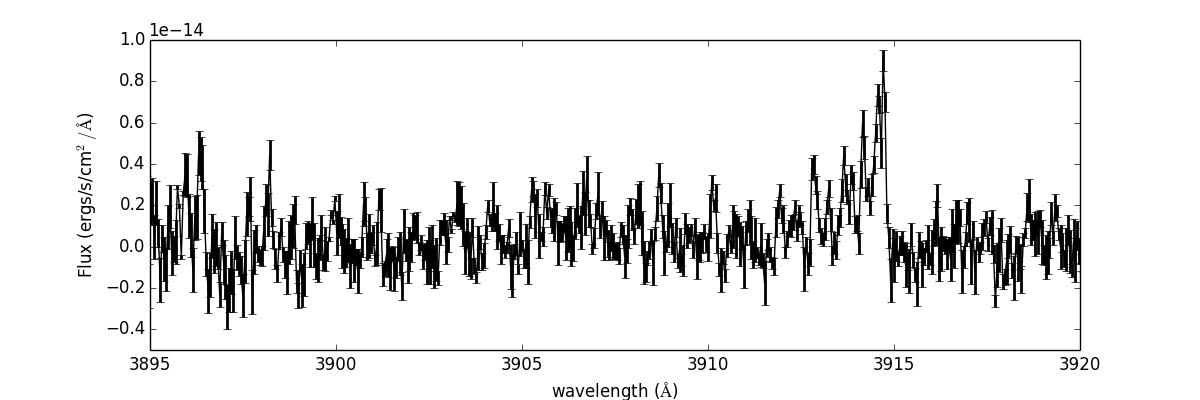}
\includegraphics[width=\textwidth]{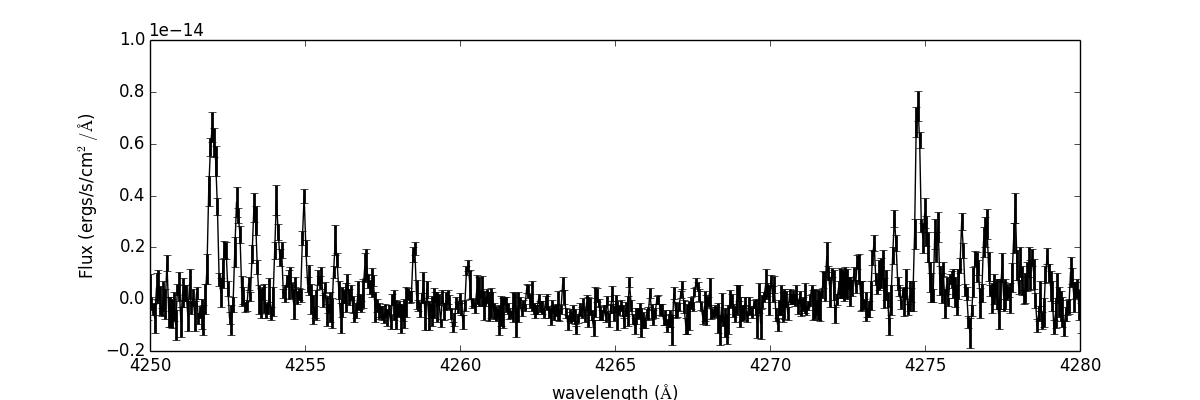}
\caption{
\label{CON2}
Spectra showing the N$_2^+$ (0-0) band (top) and CO$^+$ (2-0) band (bottom) in R2 PanSTARRS.  The derived N$_2$/CO ratio is 0.05 $\pm$ 0.01, in agreement with other measurements by~\cite{CochranMcKay2018}, ~\cite{Biver2018}, and~\cite{Opitom2019}.
}
\end{figure}

\begin{figure}[h!]
\includegraphics[width=\textwidth]{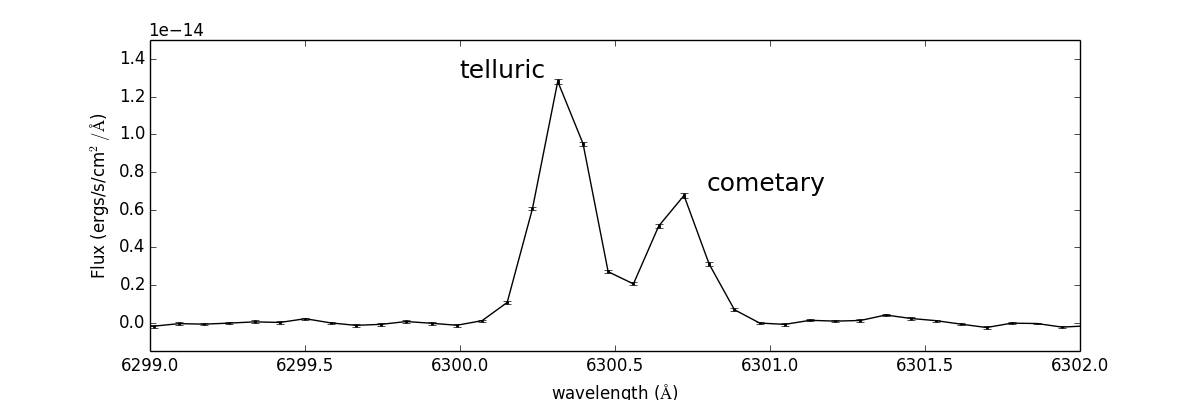}
\includegraphics[width=\textwidth]{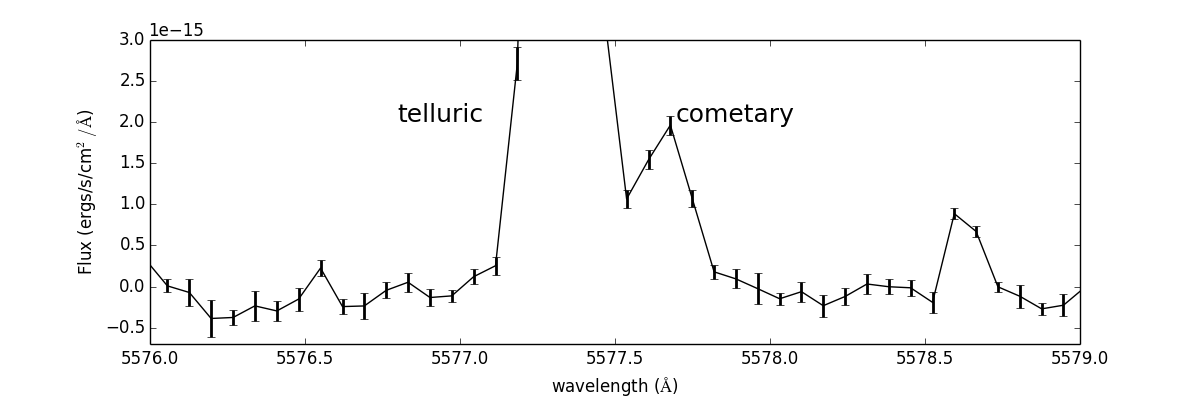}
\caption{
\label{OI}
The [\ion{O}{1}]6300~\AA~(top) and [\ion{O}{1}]5577~\AA~(bottom) lines in comet R2 PanSTARRS.  The telluric line is blueward of the cometary feature.  While often used as a proxy for H$_2$O production in comets, the large abundances of CO and CO$_2$ in R2 PanSTARRS make the derived water production rates an overestimate.  This conclusion is supported by the disagreement between the derived water production rates from the [\ion{O}{1}]6300~\AA~line and our OH observations.  Therefore we do not consider the [\ion{O}{1}]6300~\AA~line as a robust proxy for H$_2$O production in R2 PanSTARRS.}
\end{figure}

\begin{figure}[h!]
\includegraphics[width=\textwidth]{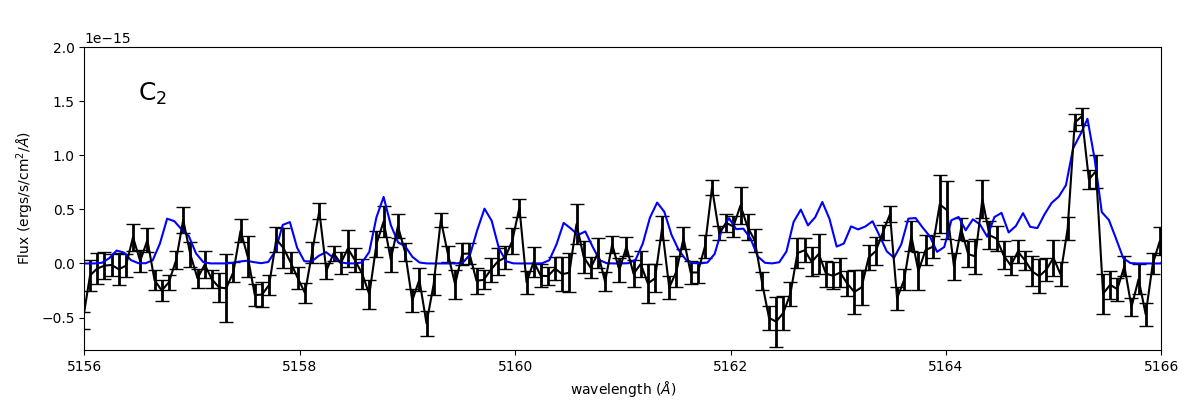}
\includegraphics[width=\textwidth]{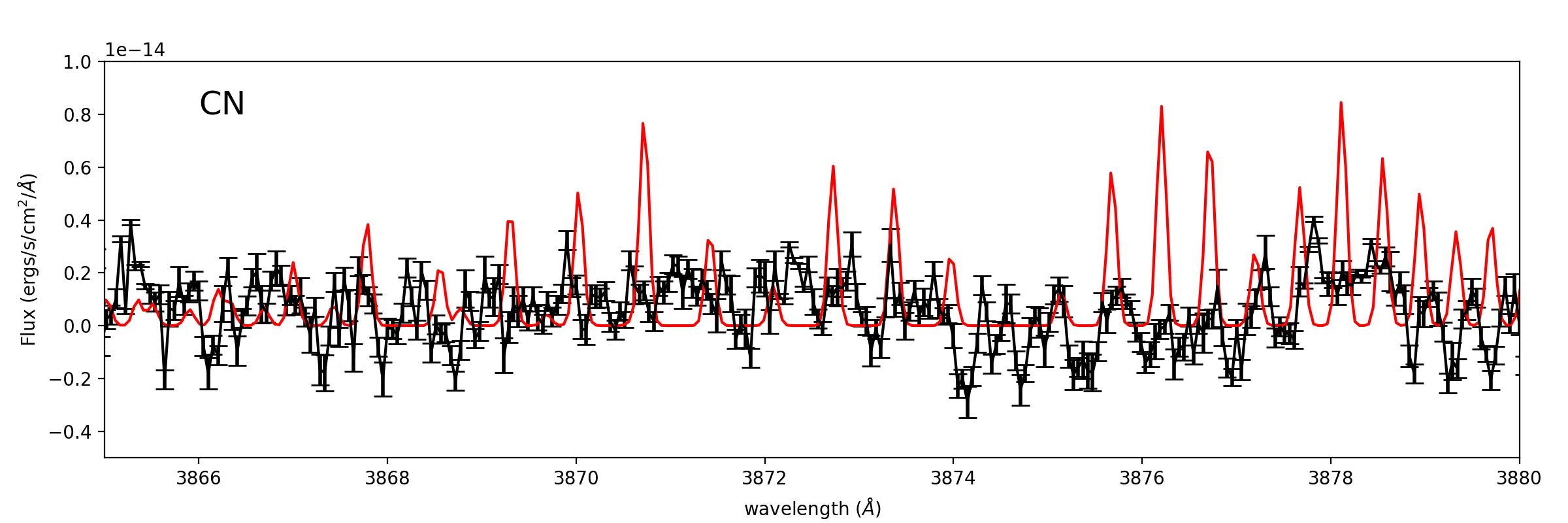}
\includegraphics[width=\textwidth]{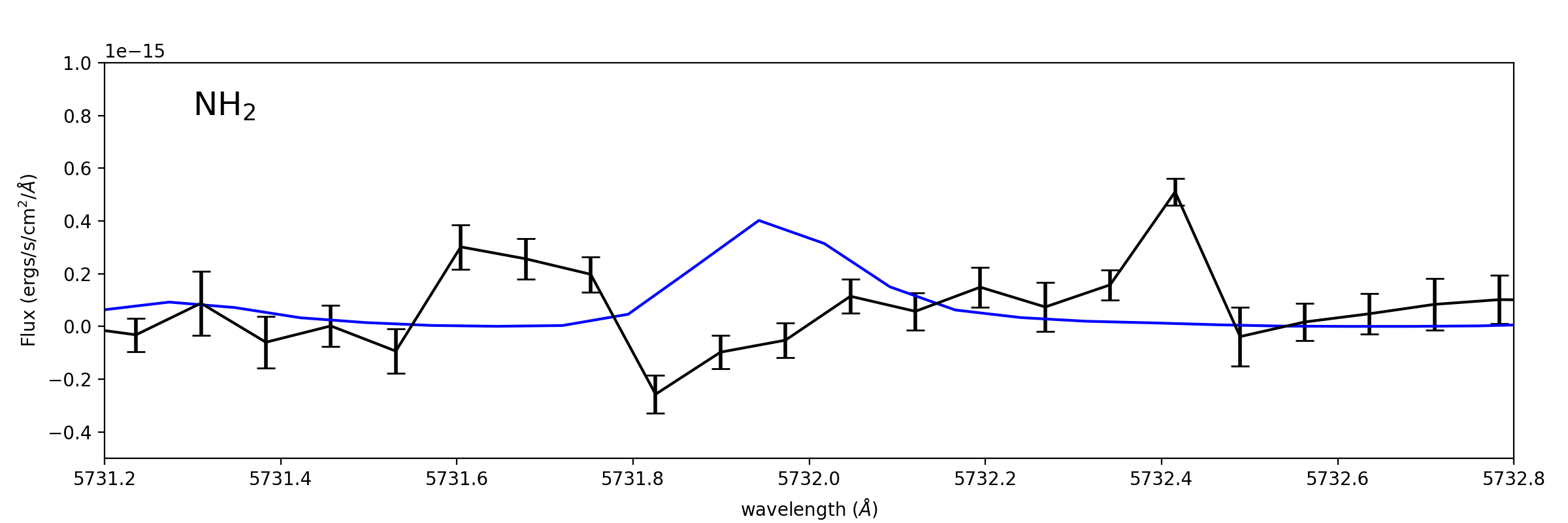}
\caption{
\label{CNNH2}
Spectral regions around C$_2$ (top), CN (middle), and NH$_2$ (bottom) emission features.  The number above the top corner of each plot (e.g. ``1e-15'' for C$_2$) denotes the multiplier for the units of the y-axis.  CN and NH$_2$ are not detected, while there is a candidate feature corresponding to the C$_2$ bandhead.  Overplotted are empirical fits to a spectrum of C/2009 P1 (Garradd) at similar heliocentric distance scaled to the flux levels of the observed R2 PanSTARRS spectra, which serve as 3$\sigma$ upper limits on the observed flux.  For C$_2$ the model matches the peak flux of the candidate bandhead at 5165~\AA.  Although we interpret this as a tentative detection of C$_2$, for this work we are employing C$_2$ as a constraint on the C$_2$H$_2$ abundance, so even a detection of C$_2$ only provides an upper limit on C$_2$H$_2$ due to multiple sources of the C$_2$ molecule in comets.  These upper limits then provide us with an upper limit on the production rate.
}
\end{figure}

\indent For the ARCES observations, we show our detections of N$_2^+$ and CO$^+$ in Fig.~\ref{CON2} and the detection of [\ion{O}{1}]6300~\AA~and [\ion{O}{1}]5577~\AA~emission in Fig.~\ref{OI}.  We do not detect CN or NH$_2$ emission and have a tentative detection of the C$_2$ bandhead at 5165~\AA, all of which we show in Fig.~\ref{CNNH2}.

\indent To derive upper limits on production rates, we employ empirical spectral fits of comet C/2009 P1 (Garradd) at a similar heliocentric distance using the spectral fitting model of~\cite{McKay2014} and scale these models so that the strongest lines would be present at the 3$\sigma$ level (for C$_2$ we scaled the model so that it corresponded to the candidate bandhead feature).  We then integrated over this model to obtain an upper limit on the observed flux and convert to production rate using our Haser model as described above.  We present the production rate upper limits for CN, NH$_2$, and C$_2$ in Table~\ref{Qrates}. Regardless of whether the inferred flux for C$_2$ is interpreted as an upper limit or a detection, the upper limit inferred for C$_2$H$_2$ is unchanged since C$_2$ has multiple sources in cometary comae~\citep{Combi1997}.  For ARCES observations we compare to the iSHELL observations of CO because the data were taken contemporaneously and both are narrow slit spectrometers sampling similar portions of the coma.\\ 
\indent From the ARCES spectra, we measure an N$_2^+$/CO$^+$ ratio of 0.05 $\pm$ 0.01.  We follow the arguments of~\cite{CochranMcKay2018} and assume that N$_2^+$/CO$^+$=N$_2$/CO (see Section~\ref{sec:discuss} for discussion of the validity of this assumption).  Therefore we infer an N$_2$/CO ratio of 0.05 $\pm$ 0.01.  Using the CO production rate derived from our iSHELL observations, we derive an N$_2$ production rate of 4.8 $\times$ 10$^{27}$ mol s$^{-1}$ (see Table~\ref{Qrates}).\\
\indent We present the H$_2$O production rates derived from the [\ion{O}{1}]6300~\AA~emission measured by ARCES and OH emission from DCT in Table~\ref{Qrates}.  There is a clear discrepancy (factor of $\sim$ 30) between the H$_2$O production rate derived from the [\ion{O}{1}]6300~\AA~emission and the value derived from our OH narrowband observations.  However, the methodology for deriving water production rates from [\ion{O}{1}]6300~\AA~emission assumes that H$_2$O photodissociation is the dominant source of \ion{O}{1} in the coma.  While this is a valid assumption for most comets and has been used in the past to derive reliable H$_2$O production rates~\citep[e.g.,][]{Morgenthaler2001, Fink2009, McKay2018}, the large CO and CO$_2$ production rates found both by ourselves and other authors compared to our derived water production rate suggest this is not a valid assumption for R2 PanSTARRS.  Not accounting for contributions from the photodissociation of other oxygen-bearing species such as CO and CO$_2$ when they are more abundant than H$_2$O results in a large overestimate of the water production rate, which is what we observe when comparing to the production rate derived from our OH observations.  As OH release into the coma is always dominated by H$_2$O, it is a more reliable proxy for H$_2$O production than [\ion{O}{1}].  Therefore we adopt the value of Q$_{H_2O}$ inferred from our OH observations as a more accurate measure of the H$_2$O production rate.\\
\indent The \ion{O}{1} photochemistry in comets has been of interest for some time~\citep[e.g.,][]{FestouFeldman1981,Cochran2008, McKay2013, McKay2015, Decock2013, Decock2015}, and the study of both the [\ion{O}{1}]6300~\AA~line as well as the [\ion{O}{1}]5577~\AA~line (which we also detect in our ARCES spectra, see Fig.~\ref{OI}) in such a CO and CO$_2$-rich comet could provide new insights into \ion{O}{1} photochemistry in comets.  We measure the flux ratio of the [\ion{O}{1}]5577~\AA~line to the sum of the [\ion{O}{1}]6300~\AA~and [\ion{O}{1}]6364~\AA~lines (the oxygen line ratio) to be 0.20 $\pm$ 0.03, in agreement with the value of 0.23 $\pm$ 0.03 measured by~\cite{Opitom2019}.  This ratio is higher than typically observed but is similar to other comets observed at heliocentric distances near 3 AU~\citep{McKay2012,Decock2013,McKay2015}.  This is consistent with the large CO$_2$/H$_2$O and CO/H$_2$O ratios observed.  However, a detailed study of the \ion{O}{1} photochemistry in R2 PanSTARRS is beyond the scope of this paper and will be pursued in a future publication.\\

\subsection{Summary of Production Rates and Relative Abundances}\label{subsec:Qrates}
\indent A summary of our resulting production rates and abundances is presented in Table~\ref{Qrates}.  We pursued several methods for determining the water production rate: H$_2$O and OH prompt emission from iSHELL, OH emission from DCT, and [OI]6300 emission from APO ARCES.  The OH data from DCT provided the best measure of H$_2$O production and therefore all mixing ratios compared to H$_2$O use the DCT-derived Q(H$_2$O) as the reference.  While we detected [\ion{O}{1}]6300 emission there is significant evidence that this emission is actually dominated by contributions from CO$_2$ and/or CO photodissociation rather than H$_2$O, and so for R2 PanSTARRS it does not serve as a reliable proxy for H$_2$O (see Section~\ref{subsec:APO}).
\begin{table}[h!]
\begin{center}
\caption{\textbf{Production Rates and Abundances}
\label{Qrates}
}
\begin{tabular}{ccccc}
\hline
Instrument & Species & Q (10$^{27}$ mol s$^{-1}$) & X/H$_2$O$^a$ (\%) & X/CO$^b$ (\%)\\
\hline
IRTF-iSHELL & CO & 95.4 $\pm$ 9.1 & (3.08 $\pm$ 0.35) $\times$ 10$^4$ & -\\
 & CH$_4$ & 0.56 $\pm$ 0.07 & 181 $\pm$ 25 & 0.59 $\pm$ 0.09\\
 & OCS & $<$ 0.23 & $<$ 74 & $<$ 0.24\\
 & C$_2$H$_6$ & $<$ 0.085 & $<$ 27 &  $<$ 0.089\\
 & CH$_3$OH & $<$ 0.36 & $<$ 116 & $<$ 0.38\\
 & H$_2$CO & $<$ 0.49 & $<$ 158 & $<$ 0.51\\
 & H$_2$O$^c$ & $<$ 78.4 & - & $<$ 82\\
 & H$_2$O$^d$ & $<$ 72.9 & - & $<$ 76\\
\hline
\textit{Spitzer} IRAC & {CO$_2$}$^e$ & 10.0 $\pm$ 1.0 & 3230 $\pm$ 380 & 18.2 $\pm$ 3.5\\
 & Af$\rho$ & 890 $\pm$ 19 cm & -23.54 $\pm$ 0.03$^f$ & -25.79 $\pm$ 0.04$^g$\\
\hline
ARO-SMT & CO & 55.0 $\pm$ 8.9 & (1.77 $\pm$ 0.31) $\times$ 10$^4$ & -  \\
\hline
APO-ARCES & N$_2$ & 4.8 $\pm$ 1.1$^h$ & 1550 $\pm$ 370 & 5.0 $\pm$ 1.0\\
 & CN & $<$ 0.02 & $<$ 6.5 & $<$ 0.021 \\
 & NH$_2$ & $<$ 0.01 & $<$ 3.2 & $<$ 0.010\\
 & C$_2$ & $<$ 0.021 & $<$ 6.8 & $<$ 0.022\\
 & H$_2$O (via [\ion{O}{1}]) & 10.5 $\pm$ 0.5 & - & 11.0 $\pm$ 1.2\\
\hline
DCT-LMI & H$_2$O (via OH) & 0.31 $\pm$ 0.02 & - & 0.32 $\pm$ 0.04\\
 & Af$\rho$ & 561 $\pm$ 6 cm & -23.74 $\pm$ 0.03$^f$ & -25.99 $\pm$ 0.04$^g$\\
\hline
\end{tabular}
\end{center}
$a$ Compared to the measured value from OH observations with the DCT, as this is the most robust measure of the H$_2$O production rate.\\
$b$ Compared to the iSHELL CO production rate except for CO$_2$ and Af$\rho$, which is compared to the SMT CO value as this value was measured contemporaneously and was used to correct the \textit{Spitzer} observations for the expected CO contribution.  N$_2$/CO is calculated from the optical observations of N$_2^+$ and CO$^+$ as described in the text.\\
$c$ Derived from H$_2$O lines in the M2 setting.\\
$d$ Derived from OH prompt emission in the Lp1 setting.\\
$e$ Derived accounting for the expected CO contribution based on the SMT observations using the methodology detailed in Section~\ref{subsec:spitzerda}.\\
$f$ $\log(Af\rho/Q_{H_2O})$\\
$g$ $\log(Af\rho/Q_{CO})$\\
$h$ Calculated using the N$_2$/CO ratio derived from optical observations multiplied by the iSHELL CO production rate.
\end{table}
\subsection{Active Areas}
We employ our CO$_2$, H$_2$O, and CO production rates to calculate the active areas of the cometary surface using the sublimation model of~\cite{CowanAHearn1979}, which are given in Table~\ref{activearea}. Following arguments given in~\cite{Bodewits2014}, ~\cite{McKay2017}, and ~\cite{McKay2018}, we adopt the slow rotator model, for which every facet of the nucleus surface is in equilibrium with the solar radiation incident upon it.  While the size of the nucleus of R2 PanSTARRS is not known, we convert these active areas to active fractions for a 1 km and 10 km radius nucleus, which encompasses the range of most cometary nuclei for which measurements are available, including an upper limit of $R < $ 15 km determined for R2 PanSTARRS from~\cite{WierzchosWomack2018}.

\begin{table}[h!]
\begin{center}
\caption{\textbf{Active Areas and Active Fractions}
\label{activearea}
}
\begin{tabular}{cccccc}
\hline
Species & Vaporization Rate & Production Rate & Active Area & Active Fraction & Active Fraction\\
 & (10$^{16}$ mol s$^{-1}$ cm$^{-2}$) & (10$^{27}$ mol s$^{-1}$) & (10$^{10}$cm$^2$) & (r=1 km, \%) & (r=10 km, \%)\\
 \hline
CO & 30.0 & 95.4 $\pm$ 9.1$^a$ & 31.8 $\pm$ 3.0 & 252 $\pm$ 24 & 2.52 $\pm$ 0.24\\
CO & 30.0 & 55.0 $\pm$ 9.0$^b$ & 18.3 $\pm$ 3.0 & 145 $\pm$ 24 & 1.45 $\pm$ 0.24\\
CO$_2$ & 8.22 & 10.0 $\pm$ 1.0 & 12.0 $\pm$ 1.2 & 96 $\pm$ 10 & 0.96 $\pm$ 0.10\\
H$_2$O & 2.00 & 0.31 $\pm$ 0.02 & 1.6 $\pm$ 0.1 & 12.2 $\pm$ 0.8 & 0.122 $\pm$ 0.008\\
\end{tabular}
\end{center}
$a$ iSHELL-derived CO production rate\\
$b$ SMT-derived CO production rate
\end{table}
\section{Discussion} \label{sec:discuss}
\subsection{Other Potential Molecular Emissions in \textit{Spitzer} Imaging}
\indent As our CO$_2$ abundance is dependent on an analysis of broadband imaging, not spectroscopy, it is possible there are other emission features in our bandpass besides CO$_2$ and CO, the species commonly assumed to dominate the flux.  This would lead to an overestimate of the CO$_2$ production rate.  The leading neutral candidates are OCS and N$_2$O.  Hot bands of water in the 4.5-5.0$\mu$m region would normally be important \citep{Ootsubo2012} but the extremely low H$_2$O/CO and H$_2$O/CO$_2$ ratios for R2 PanSTARRS mean they can be neglected.\\
\indent OCS has a strong band at $\sim$ 4.9 $\mu$m, and has been observed in comets at an abundance of approximately 0.1-0.4\% with respect to H$_2$O~\citep[][and references therein]{BockeleeMorvan2004, MummaCharnley2011, DelloRusso2016}.  OCS is covered by the M2 grating setting in our iSHELL observations (Fig.~\ref{iSHELL_stack}), and was not detected, placing an upper limit on its abundance ratio compared to CO of 0.24\%.  At this level OCS is not a significant contributor to the 4.5 $\mu$m band flux.\\
\indent The other neutral molecule with strong emissions in the 4.5$\mu$m bandpass is N$_2$O.  To date, N$_2$O has never been detected in a comet, despite ISO and AKARI observations that cover this wavelength range~\citep{Crovisier1997b, Ootsubo2012}.  As of this writing we are not aware of any reported detection of N$_2$O by the Rosetta instruments around 67P/Churyumov-Gerasimenko.  However, the large N$_2$ abundance and the presence of an intrinsically strong N$_2$O band at $\sim$4.5 $\mu$m could perhaps indicate a strong contribution of this molecule to our \textit{Spitzer} photometry.  The N$_2$O emission is too far to the blue to be covered by our iSHELL observations in the M2 grating, and also the region around the N$_2$O emission band is heavily affected by telluric CO$_2$ absorption.  Therefore we cannot definitively rule out the presence of N$_2$O emission in our \textit{Spitzer} images.  The g-factor for the relevant N$_2$O transition is 1.5 $\times$ 10$^{-3}$ photons/s/mol, meaning that an N$_2$O production rate of $\sim$1.8 $\times$ 10$^{28}$ mol/s could account for the residual flux we attribute to CO$_2$.  This suggests that even at $\sim$1.8 $\times$ 10$^{27}$ mol/s N$_2$O could account for 10\% of the flux we attribute to CO$_2$.  However, N$_2$O is also very efficient at releasing O($^1$S) into the coma~\citep{Huebner1992}, and at these high production rates we may expect it to dominate the \ion{O}{1} photochemistry and drive the oxygen line ratio to be $\sim$1.  However, we measure an oxygen line ratio of 0.20 (Section 3.5), suggesting that N$_2$O is not a dominant source of \ion{O}{1} in the coma.  While \ion{O}{1} photochemistry in comets is not well understood, this, combined with a lack of previous detections in comets, likely implies that N$_2$O is not contributing much more than 10\% to the residual flux we attribute to CO$_2$.\\
\indent Lastly, the $\Delta$v=1-0 band of CO$^+$ is located at $\sim$4.6 $\mu$m.  We do indeed observe a faint tail feature with structures reminiscent of an ion tail in both \textit{Spitzer} epochs (Fig.~\ref{Spitzer}).  We attribute this to the CO$^+$ band based on the large CO production rate and the strong CO$^+$ tails observed in the optical for R2 PanSTARRS.  However, this tail is much fainter than the neutral emission and only appears obvious in Figure~\ref{Spitzer} due to the image enhancement process.  Additionally, this band is covered by our iSHELL observations yet is not detected.  Therefore we conclude that CO$^+$ cannot account for the observed excess emission that we attribute to CO$_2$.\\
\indent While we cannot unequivocally rule out N$_2$O emission, we conclude that our \textit{Spitzer} 4.5 $\mu$m band imaging is most likely dominated by a combination of CO$_2$ and CO emission.  After accounting for the expected CO contribution based on independent observations with other facilities, we think that our analysis provides an accurate measure of the CO$_2$ abundance in R2 PanSTARRS.
\subsection{Comparison to Other Observations of R2 PanSTARRS}
\indent Our N$_2$/CO ratio agrees well with the measurements made by~\cite{CochranMcKay2018} in December 2017, almost two months before our observations, as well as measurements by~\cite{Biver2018} made about a week before our observations and measurements by~\cite{Opitom2019} about two weeks after our observations.  This suggests there was little evolution in the N$_2$/CO ratio over the December 2017-February 2018 time frame.\\
\indent Our iSHELL CO production rate agrees well with values determined by~\cite{deValBorro2018} and \cite{Biver2018}, whose observations were 1-2 weeks before ours (although these works used an asymmetric outgassing model while we assume symmetric outflow).  Our value is higher than that found by~\cite{WierzchosWomack2018} using the SMT in late December 2017 and mid-January 2018, as well as the value we measure from our SMT observations.  This could be due to different FOV or the narrow nature of our slit.  There is evidence from the millimeter observations of~\cite{Biver2018} and~\cite{WierzchosWomack2018} that there is a significant sunward outgassing component, an enhancement also observed in our \textit{Spitzer} imaging (see Fig.~\ref{Spitzer}).  As the iSHELL slit was oriented along the Sun-comet line, we may have sampled this enhancement, which coupled with our application of a symmetric outgassing model may result in an overestimate of the CO production rate, but we do not observe a strong asymmetry in the CO line profile (see Section~\ref{subsec:ishellda}, Fig.~\ref{Qcurve}), meaning the iSHELL observations were not sensitive to the asymmetry inferred from the mm and \textit{Spitzer} observations.  One reason for this discrepancy could be the limited spatial coverage of the iSHELL slit (slit width 0.75\arcsec $\times$ length $\sim$6\arcsec) compared to the much larger circular beams in the sub-mm (10-30\arcsec) and the arcminute spatial scale covered by the \textit{Spitzer} imaging.  A narrow feature that is only slightly offset from the solar direction could be missed by the narrow iSHELL slit, while still revealing an asymmetry that would be evident only on larger spatial scales.  There is evidence from the coma morphology in the \textit{Spitzer} images that the asymmetry manifests as a large fan-shaped feature (Fig.~\ref{Spitzer}, rightmost column), perhaps implying the asymmetry would not be as pronounced in the inner coma sampled by iSHELL.\\ 
\indent In analyzing our iSHELL spectra we adopt the gas expansion velocity (v$_{exp}$) measured by SMT (0.52 km/s), but as the iSHELL slit samples the very inner coma, it is possible the gas may not have been fully accelerated to the speed observed at larger nucleocentric distances.  A $\sim$ 30\% smaller assumed v$_{exp}$ (i.e., from 0.52 km/s to 0.35-0.40 km/s) would bring the iSHELL and SMT results for $Q_{CO}$ into agreement within their respective 1-$\sigma$ uncertainties.  As the resolving power of our iSHELL spectra ($\lambda / \Delta \lambda \sim 4 \times 10^4$) is insufficient to resolve the cometary lines, we obtain no information regarding v$_{exp}$ and therefore we cannot confirm or refute the possibility of a lower expansion velocity in the inner coma.  However, the rotational temperature we measured for CO (13$\pm$2 K) is significantly lower than that obtained for CH$_3$OH from IRAM of 23K~\citep{Biver2018}.  This discrepancy may be due to photolytic heating of the coma as the gas expands.  Based on this result we might expect a smaller expansion velocity to be more relevant to the iSHELL observations since the terminal region spanned lines-of-sight offset only $\sim$1-5\arcsec~from the nucleus, in contrast to the much larger SMT/IRAM beam sizes ($\sim$ 10-30\arcsec).  In any case, while absolute production rates are proportional to v$_{exp}$ (assuming the beam size is small compared to the photodissociation scale length, which is the case for all species reported here), relative abundances (i.e., mixing ratios) are relatively insensitive to the adopted value of v$_{exp}$ as long as a common v$_{exp}$ is assumed for all species, which is the case in this work.  Therefore, we would expect negligible changes in abundance ratios for the other species included in the M2 and Lp1 settings (e.g., Table 3).\\
\indent The different rotational temperature measured from our iSHELL observations versus the IRAM observations of~\cite{Biver2018} and adopted for our SMT analysis could affect our results.  For the iSHELL observations adopting $T_{rot}$=23K results in a $\sim$10\% higher CO production rate, while adopting $T_{rot}$=13K for the SMT observations results in a CO production rate $\sim$30\% higher than the value for $T_{rot}$=23K.  Adopting the higher $T_{rot}$ for the iSHELL data results in a $\sim$70\% larger CH$_4$ production rate, 15-20\% less sensitive upper limits on CH$_3$OH and C$_2$H$_6$, a 15\% more sensitive upper limit on H$_2$CO, and a 10\% more sensitive upper limit on OCS.  In terms of mixing ratios compared to CO, the CH$_4$/CO ratio is most affected, having a value 50\% higher for $T_{rot}$=23K than for $T_{rot}$=13K.  The upper limits on C$_2$H$_6$/CO and CH$_3$OH/CO change minimally ($\sim$ 5\%) for the different value of $T_{rot}$, while the upper limits on H$_2$CO/CO and OCS/CO are approximately 20\% more sensitive for $T_{rot}$=23K.  Analysis for the other instruments does not depend on $T_{rot}$, though the slightly higher CO production rates determined from both iSHELL and SMT data using the alternate rotational temperature would result in proportionately lower values of the mixing ratios compared to CO.\\
\indent Our upper limit for CH$_3$OH is a factor of three lower than the detection by~\cite{Biver2018} (this is independent of the value of $T_{rot}$, as are all the comparisons in this paragraph), which could be due to temporal and/or spatial variability in CH$_3$OH outgassing.  The different projected areas observed for the IRAM and iSHELL observations could account for this discrepancy if a fraction of CH$_3$OH is released as icy grains in the coma.  At the same time, our H$_2$CO upper limit is consistent with the detection of~\cite{Biver2018}.  Our HCN upper limit (inferred from CN) is consistent with both the upper limit found by~\cite{WierzchosWomack2018} and the detection reported by~\cite{Biver2018}.  Our H$_2$O production rate inferred from the OH DCT data is consistent with the upper limits reported by~\cite{Biver2018} from observations of the OH 18 cm line at Nancay and by~\cite{Opitom2019} using narrowband photometry from the TRAPPIST telescope.\\  
\indent At the time of this writing our results are the first reported detections or upper limits on CO$_2$, CH$_4$, C$_2$H$_6$, C$_2$H$_2$, OCS, and NH$_3$ in R2 PanSTARRS.  While~\cite{Opitom2019} report a lower limit on the N$_2^+$/NH$_2$ ratio (a proxy for N$_2$/NH$_3$), this is a lower limit on the column density ratio in the slit, and because of the different spatial distributions of ions and neutrals, interpreting the column density ratio as an actual abundance ratio is complicated.  Assuming N$_2^+$/CO$^+$=N$_2$/CO, we have used our neutral CO measurements to convert the N$_2^+$/CO$^+$ to an equivalent N$_2$ production rate, which then can be compared to the production rates derived for other neutrals such as NH$_3$ and HCN, as was also done by \citet{WierzchosWomack2018}.  With those caveats aside, the lower limit on N$_2$/NH$_3$ provided by~\cite{Opitom2019} is consistent with our constraints.\\
\indent ~\cite{Opitom2019} also detected CO$_2^+$ emission and derive a CO$_2^+$/CO$^+$ ratio of 1.1 $\pm$ 0.3.  They do not interpret this as the CO$_2$/CO ratio, as CO$_2$ can also contribute to observed CO$^+$ emission.  For this reason~\cite{Opitom2019} also express caution when interpreting the N$_2^+$/CO$^+$ ratio as a direct measurement of N$_2$/CO, as CO$_2$ photodissociation could contribute to the observed CO$^+$ emission and therefore the measured N$_2^+$/CO$^+$ ratio actually only provides a lower limit on N$_2$/CO.  However, from our \textit{Spitzer} observations of neutral CO$_2$, we derive CO$_2$/CO $\sim$ 18\%, and at this low abundance CO photoionization should be the dominant source of CO$^+$ ions~\citep{Huebner1992}.  This appears to disagree with the large CO$_2^+$/CO$^+$ ratio observed by~\cite{Opitom2019}.  The reason for the discrepancy between ion and neutral observations is unknown, but may be related to our understanding of CO$_2$ and CO photochemistry.\\
\indent The Af$\rho$ value we find for R2 PanSTARRS from our DCT BC imaging is lower than other measurements by~\cite{Opitom2019} as well as from the cometary database developed by T. Noel\footnote{http://www.lesia.obspm.fr/comets/} quoted by~\cite{Biver2018}.  There is some evidence for temporal variability in the observations of~\cite{Opitom2019}, so that may explain some of the discrepancy.  Variable gas contamination by strong CO$^+$ emission could also be present, especially for the broadband photometry quoted by~\cite{Biver2018}.  The Af$\rho$ value derived from our \textit{Spitzer} imaging is higher than all values reported in the optical, including our DCT observations, which were contemporaneous with the Spitzer epoch on UT February 21.  However, the different wavelength regimes for the IR and optical data make it unclear how comparable the Af$\rho$ values actually are.  For the \textit{Spitzer} data, we employed a larger photometric aperture than the optical studies (28,000 km vs. 10,000 km); using a smaller aperture in line with other studies yields Af$\rho$ $\sim$ 1000 cm, larger than found for the original aperture.  While Af$\rho$ is independent of aperture size for ideal comae, the presence of a dust tail and acceleration of the dust can account for the decreasing trend with aperture size we observe for Af$\rho$.  Spectral reddening of light scattered by comae can vary with wavelength, especially over such a large spectral range \citep{Jewitt1986}.  Nevertheless, we calculate the effective BC (0.45~\micron) to 3.6~\micron{} spectral slope as $\sim$1.8\% per 100~nm, similar to other comets observed in the NIR by~\cite{Jewitt1986}.
\subsection{Comparison to Other Comets}
\indent In this section we compare R2 PanSTARRS to other comets.  First we compare to the cometary population as a whole, then specifically to other comets observed at similar heliocentric distance.
\subsubsection{Cometary Population}
\indent Table~\ref{abundances} summarizes derived production rates (or upper limits) for R2 PanSTARRS, along with mixing ratios relative to H$_2$O and CO, for 12 species.  For comparison, the average values of these mixing ratios for the sample of Oort Cloud Comets observed to date are also presented.  Most values for R2 PanSTARRS are from this work, though for CH$_3$OH, HCN, and H$_2$CO our observations only provide upper limits while~\cite{Biver2018} secured detections, so in Table~\ref{abundances} we present the detected production rates for these species from~\cite{Biver2018}.\\
\indent All detected species in R2 PanSTARRS are heavily enriched compared to H$_2$O.  CO is enriched by four orders of magnitude with N$_2$ enriched by a similar amount (though the small sample size of N$_2$ measurements in comets makes this difficult to quantify), while CH$_4$, CH$_3$OH, and CO$_2$ are enriched by a factor of $\sim$ 160-200.  H$_2$CO and HCN have less drastic enrichments of $\sim$ 40 and 6, respectively, though these are still extraordinary enhancements never before observed in a comet.  We cannot draw any conclusions regarding the abundances of C$_2$H$_6$, NH$_3$, C$_2$H$_2$, or OCS compared to H$_2$O, though our results imply that while they could be heavily enriched compared to H$_2$O, this enrichment is not as drastic as inferred for species such as CO, and may be more in line with that observed for species such as H$_2$CO or HCN.\\
\indent While a definitive comparison to H$_2$O is not possible for all species, the strong CO detection allows comparison of all species to CO.  We find all species searched for are heavily depleted compared to CO, except for N$_2$ which is enhanced.  Even for species that aren't detected, the upper limits are sensitive enough to demonstrate heavy depletions.  All depleted species are underabundant by at least 1-3 orders of magnitude compared to other comets.\\
\indent R2 PanSTARRS has provided a glimpse into a rarely observed (alternative) compositional taxonomy, with CO replacing H$_2$O as the dominant gas in the coma. The measured abundance ratio CH$_4$/CO ($\sim$ 0.6\%; Table 3) approaches the mean CH$_4$/H$_2$O and C$_2$H$_6$/H$_2$O ratios among comets from the Oort cloud (\cite{DelloRusso2016}, \cite{BockeleeMorvan2004}). In contrast, C$_2$H$_6$ is strongly depleted (by a factor of at least 6.6 relative to CH$_4$), with C$_2$H$_6$/CO rivaling or surpassing the level of depletion of C$_2$H$_6$ (relative to H$_2$O) measured for disrupted comet C/1999 S4 (LINEAR), a current ``end-member" in terms of its severe depletion in all reported volatiles with the exception of HCN~\citep{Mumma2001}.  Another peculiarity of R2 PanSTARRS is the large N$_2$ abundance, with N$_2$ being the dominant reservoir of volatile nitrogen, more abundant than even H$_2$O (only CO$_2$ and CO are more abundant than N$_2$).  Typically NH$_3$, and to a lesser extent HCN, are the most abundant nitrogen-bearing volatiles in comets.  However, in R2 PanSTARRS NH$_3$/N$_2$ $<$ 0.21\% and HCN/N$_2$=0.08 $\pm$ 0.03\%.  So unless another more complicated form of volatile nitrogen that is not constrained by our observations (e.g., N$_2$O, C$_2$N$_2$, CH$_3$CN) is present at significant levels, more than 99\% of the volatile nitrogen in R2 PanSTARRS is contained in N$_2$.  There are not many measurements of the NH$_3$/N$_2$ value in comets; however, this limit for R2 PanSTARRS is much lower than measured for comet Halley ($\sim$1000\%), and is closer to derived values for several dense molecular clouds in star-forming regions ($\sim$0.6\%) \citep{Womack1992N2NH3}.  The very low derived relative abundance of NH$_3$/N$_2$ is consistent with the suggestion by \citet{WierzchosWomack2018} that R2 PanSTARRS formed in an environment with decreased photodissociation of N$_2$, leading to preserving, or shielding, of N$_2$ and inhibited production pathways of hydrogen-rich species, such as HCN and NH$_3$ \citep{HilyBlant2017}. 

\begin{table}[h!]
\begin{center}
\scriptsize
\caption{\textbf{Summary of Production Rates and Relative Abundances}
\label{abundances}
}
\begin{tabular}{lcccccc}
\hline
Instrument & Species & Q (10$^{27}$ mol s$^{-1}$) & X/H$_2$O (\%) & Mean X/H$_2$O$^a$ (\%) & X/CO (\%) & Mean X/CO$^b$ (\%)\\
\hline
\textit{Spitzer} & CO$_2$ & 10.0 $\pm$ 1.0 & 3230 $\pm$ 380 & 17.0 $\pm$ 6.0 & 18.2 $\pm$ 3.5 & 425 $\pm$ 178\\
\hline
IRTF iSHELL & CO & 95.4 $\pm$ 9.1 & (3.08 $\pm$ 0.35) $\times$ 10$^4$ & 4.0 $\pm$ 0.9 & - & -\\
& CH$_4$ & 0.56 $\pm$ 0.07 & 181 $\pm$ 25 & 0.88 $\pm$ 0.10 & 0.59 $\pm$ 0.09 & 22.0 $\pm$ 5.5\\
& C$_2$H$_6$ & $<$ 0.085 & $<$ 27 & 0.63 $\pm$ 0.10 & $<$ 0.089 & 15.8 $\pm$ 4.4\\
& OCS & $<$ 0.23 & $<$ 74 & $\sim$ 0.25 & $<$ 0.24 & $\sim$ 6.3\\
\hline
APO ARCES & N$_2$$^c$ & 4.8 $\pm$ 1.1 & 1550 $\pm$ 370 & $<<$ 1?$^d$ & 5.0 $\pm$ 1.0 & $<$ 1?$^d$\\
& NH$_3$$^e$ & $<$ 0.01 & $<$ 3.2 & 0.91 $\pm$ 0.30 & $<$ 0.010 & 22.8 $\pm$ 9.1\\
& C$_2$H$_2$$^f$ & $<$ 0.021 & $<$ 6.8 & 0.16 $\pm$ 0.03 & $<$ 0.022 & 4.0 $\pm$ 1.2\\
\hline
DCT LMI & H$_2$O$^g$ & 0.31 $\pm$ 0.02 & - & - & 0.32 $\pm$ 0.04 & 2500 $\pm$ 560\\
\hline
IRAM$^h$ & CH$_3$OH & 1.12 $\pm$ 0.07 & 360 $\pm$ 32 & 2.21 $\pm$ 0.24 & 1.04$\pm$ 0.08 & 55.3 $\pm$ 13.8\\
& H$_2$CO & 0.045 $\pm$ 0.007 & 14.5 $\pm$ 2.4 & 0.33 $\pm$ 0.08 & 0.043 $\pm$ 0.006 & 8.0 $\pm$ 2.6\\
& HCN & (4.0 $\pm$ 1.0) $\times$ 10$^{-3}$ & 1.3 $\pm$ 0.3 & 0.22 $\pm$ 0.03 & (3.8 $\pm$ 1.0) $\times$ 10$^{-3}$ & 5.5 $\pm$ 1.4\\
\end{tabular}
\end{center}
$a$ Mean mixing ratio compared to H$_2$O in the sample of Oort Cloud comets observed to date.  The uncertainties reflect the standard deviation in measured values. All values except for N$_2$ and CO$_2$ are from~\cite{DelloRusso2016b}.  The CO$_2$ value is from~\cite{Ootsubo2012}.\\
$b$ Mean mixing ratio compared to CO in the sample of Oort Cloud comets observed to date.  The uncertainties reflect the standard deviation in measured values.  References for abundances are the same as for H$_2$O detailed in footnote a.\\
$c$ Derived by multiplying the derived N$_2$/CO ratio from our ARCES observations by the CO production rate determined from our iSHELL observations.\\
$d$ Due to the lack of observations of N$_2$ in comets, the mean mixing ratios compared to H$_2$O and CO are not meaningful to calculate.  Therefore the number included is based on past observations discussed in~\cite{CochranMcKay2018} and has very large uncertainty (indicated by the ``?'').\\
$e$ Derived from our NH$_2$ upper limit assuming all NH$_2$ is released via NH$_3$ photodissociation (i.e. Q$_{NH_3}$=Q$_{NH_2}$).\\
$f$ Derived from our C$_2$ upper limit assuming all C$_2$ is released via C$_2$ photodissociation (i.e. Q$_{C_2}$=Q$_{C_2H_2}$).\\
$g$ Derived from our OH narrowband observations.\\
$h$ Production rates from~\cite{Biver2018}.  Mixing ratios compared to CO are taken directly from~\cite{Biver2018}, mixing ratios for H$_2$O are calculated using our derived H$_2$O production rate.
\end{table}

\begin{figure}[h!]
\includegraphics[width=0.5\textwidth]{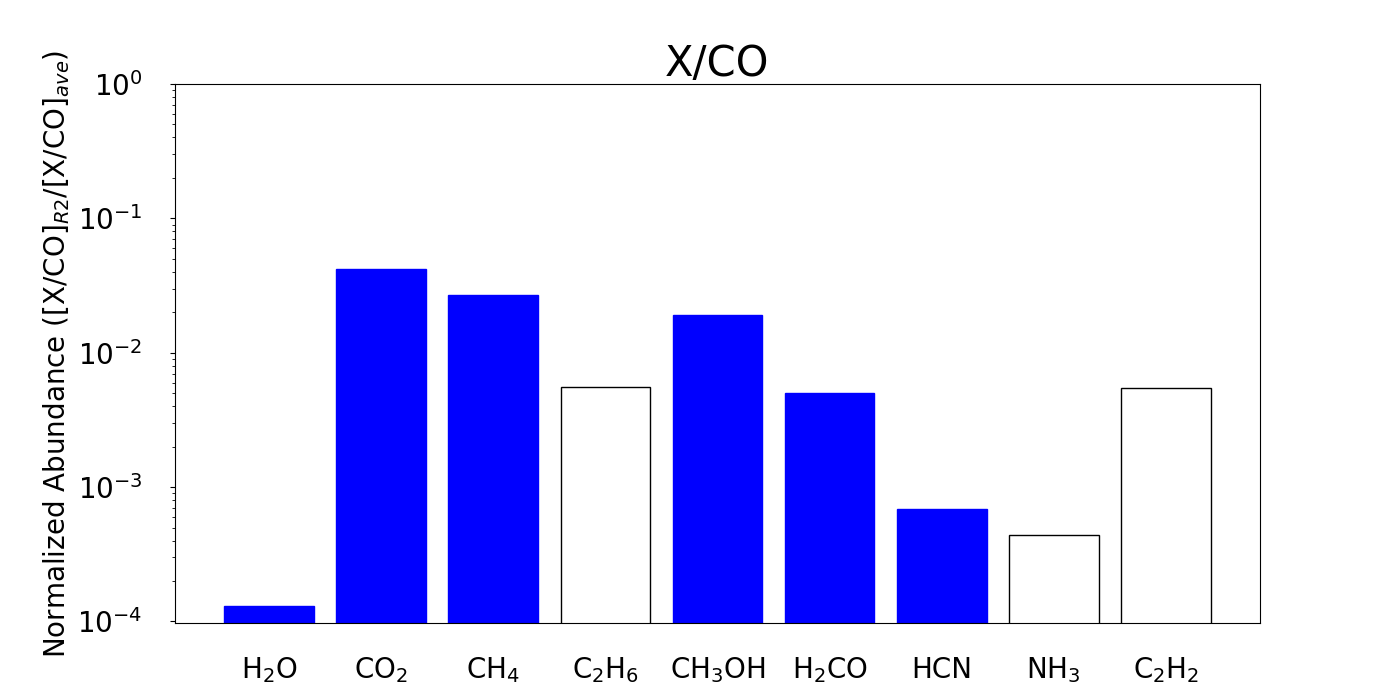}
\includegraphics[width=0.5\textwidth]{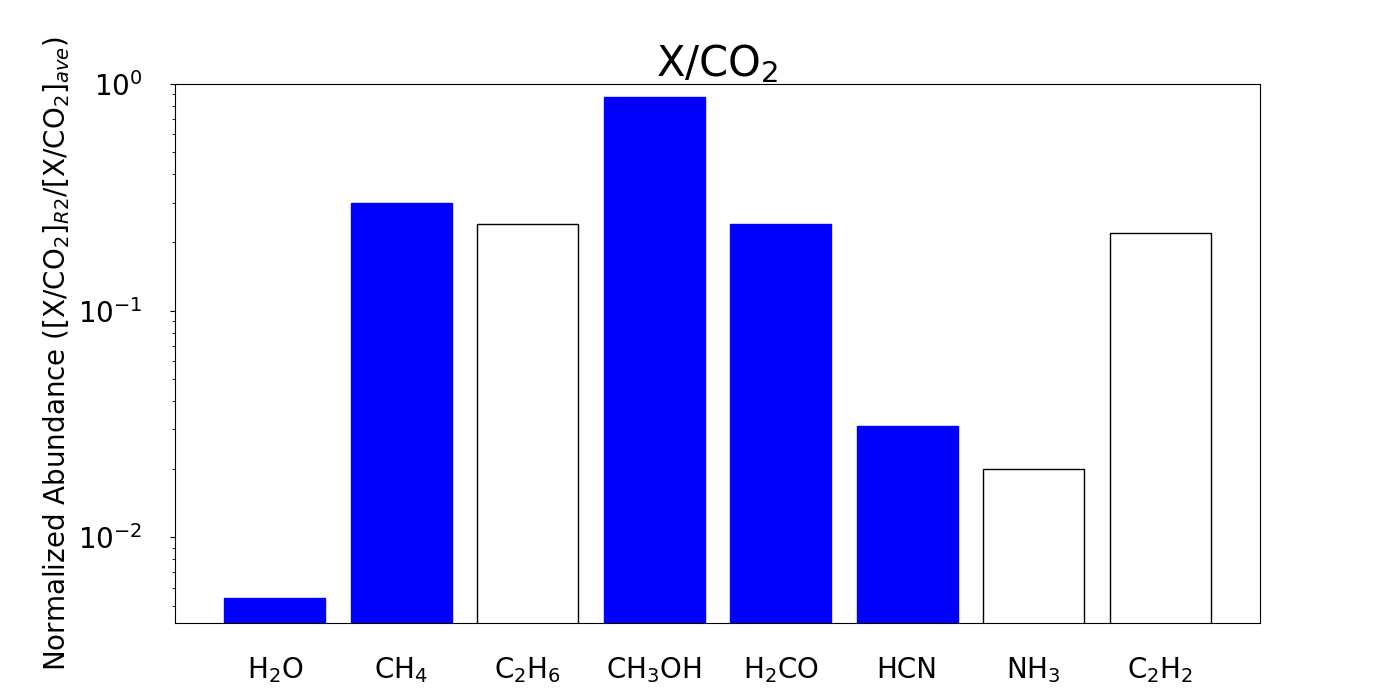}\\
\includegraphics[width=0.5\textwidth]{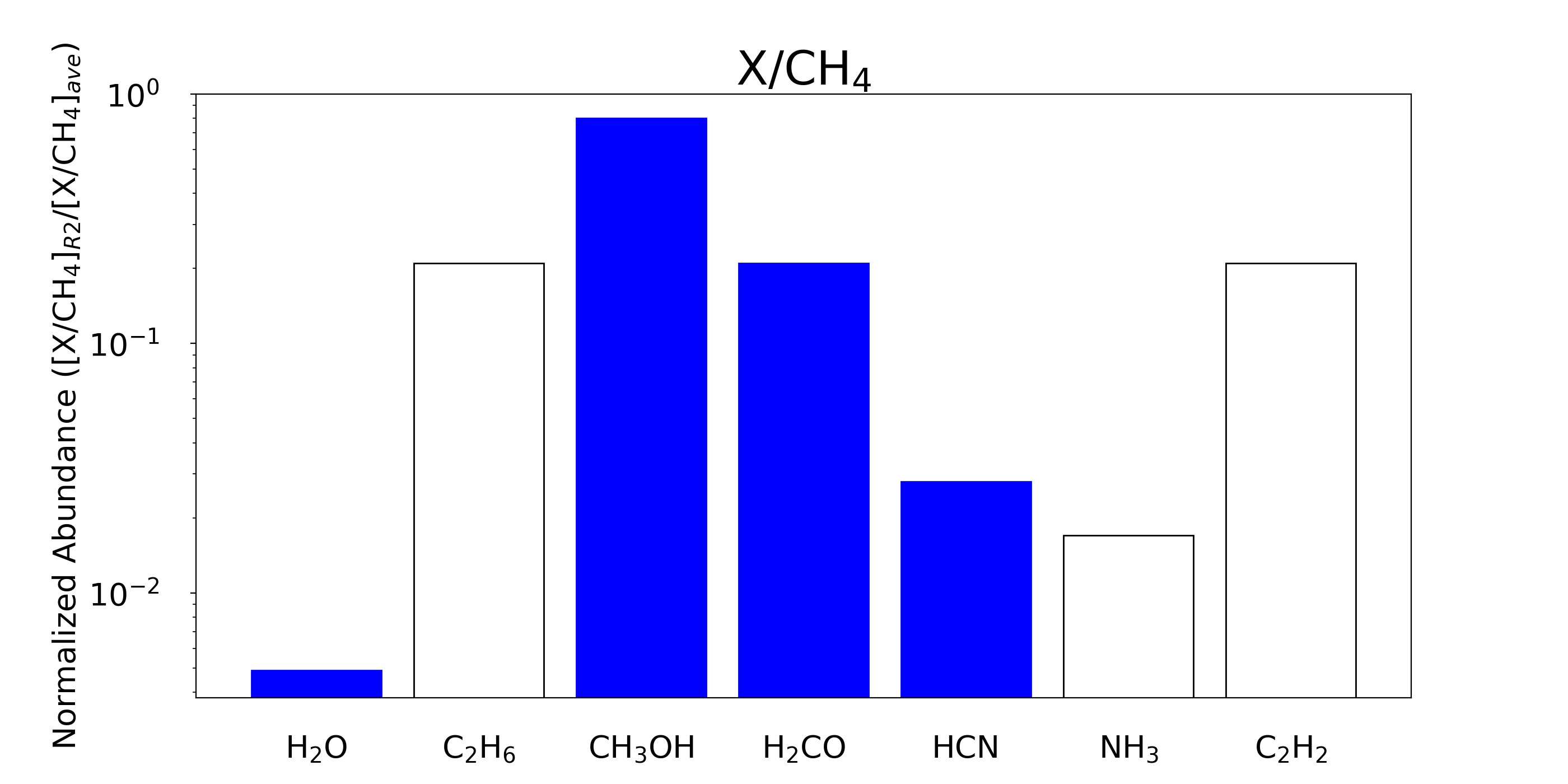}
\includegraphics[width=0.5\textwidth]{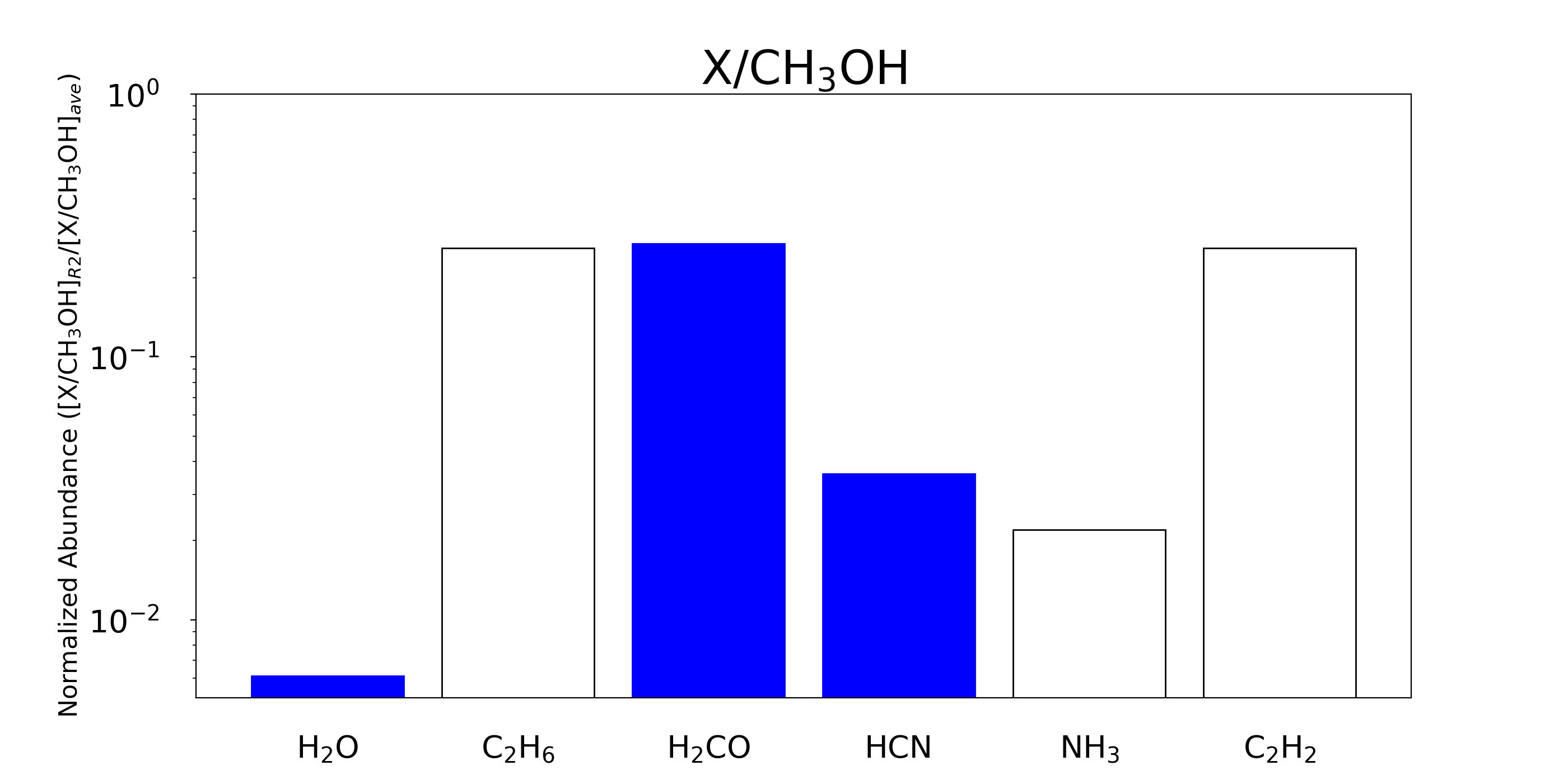}\\
\caption{
\label{compare}
Comparison of the volatile composition of C/2016 R2 (PanSTARRS) to the average comet for mixing ratios relative to four of the five most abundant ices in R2 PanSTARRS: CO (top left), CO$_2$ (top right), CH$_4$ (bottom left), and CH$_3$OH (bottom right).  The upper limit of each plot is unity, indicating a ``typical'' composition compared to other comets observed to date.  A white histogram with black outline denotes an upper limit, while blue indicates a detected value.  It is clear that no matter what species is used as the standard the composition of R2 PanSTARRS deviates from the average comet by at least a factor of three for almost all species whose abundances are constrained with sufficient sensitivity.  The exceptions are CH$_3$OH/CO$_2$ and CH$_3$OH/CH$_4$, which have values only 10-20\% below the average values for the ensemble of comets observed to date.  OCS and N$_2$ are omitted due to a lack of sufficiently well defined average values for the cometary population, but the large N$_2$ abundance (third most abundant volatile detected) certainly suggests that N$_2$ is enriched relative to all species when compared to other comets.}
\end{figure}

\begin{figure}[h!]
\includegraphics[width=\textwidth]{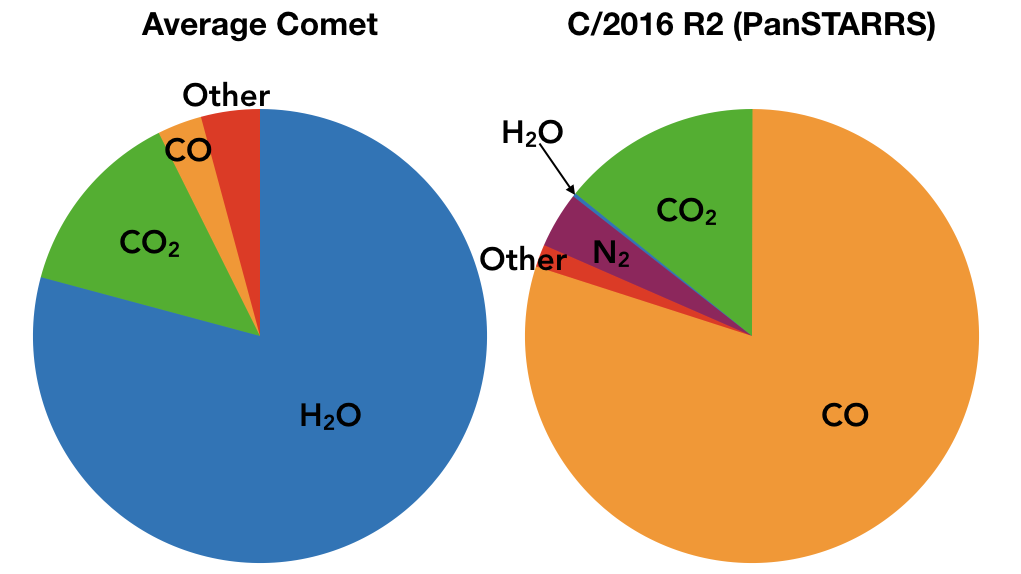}
\caption{
\label{R2_comparison}
Comparison of the volatile composition of C/2016 R2 (PanSTARRS) to the average comet as derived using mixing ratios from~\cite{DelloRusso2016b} and~\cite{Ootsubo2012}.  The dominance of CO and large amount of N$_2$ makes R2 PanSTARRS unique among comets observed with modern capabilities inside the water ice line.}
\end{figure}

\indent We show additional mixing ratios compared to CO$_2$, CH$_4$, CH$_3$OH, H$_2$CO, and HCN in Table~\ref{otherabundances}, with the mean ratio observed among comets listed in parentheses.  A subset of this compilation is shown visually as histograms in Fig.~\ref{compare}.  For mixing ratios compared to CO$_2$ we employ the average CO$_2$/H$_2$O ratio in the AKARI sample~\citep{Ootsubo2012}, while other species like CH$_3$OH have their average abundance compared to H$_2$O derived from ground-based IR studies~\citep{DelloRusso2016}, meaning that CO$_2$ and the other species were not observed contemporaneously and in most cases the sample of AKARI and ground-based IR observations sampled different comets.  This means interpretation of the average CO$_2$ abundance in Table~\ref{otherabundances} is limited by the degree to which the AKARI and ground-based IR studies provide a random sample of the cometary population.\\
\indent Almost all mixing ratios in R2 PanSTARRS deviate from typical values by at least a factor of three, although for some species that were not detected we do not have sufficient sensitivity to rule out a normal abundance (e.g. C$_2$H$_6$/HCN).  HCN is universally depleted by at least an order of magnitude compared to all detected species except H$_2$O.  Interestingly CH$_3$OH/CO$_2$ and CH$_3$OH/CH$_4$ are the only mixing ratios that are similar (within a factor of two) to the average values observed among comets.  Perhaps this commonality between the peculiar R2 PanSTARRS and other comets can shed light on its origin.  In any case, the almost universal peculiarity of the observed abundances in R2 PanSTARRS compared to other comets implies that this is not a case of one or two species being anomalous (e.g., CO being heavily enriched and H$_2$O being heavily depleted), but a complete composition fundamentally different from the ensemble of comets observed to date.\\
\indent This strong deviation from a ``typical'' cometary composition is also illustrated in Fig.~\ref{R2_comparison}.  Over 80\% of the volatile composition of R2 PanSTARRS is CO, while H$_2$O is relegated to the status of a trace volatile.  N$_2$ is also much more abundant than in typical comets, and more abundant than the other typical trace species (i.e., other than CO, H$_2$O, and CO$_2$) combined.  CO$_2$ as a fraction of the volatile inventory of R2 PanSTARRS is actually fairly close to typical comets.\\
\indent R2 PanSTARRS has a much lower Af$\rho$ than other comets with such high gas production observed at similar heliocentric distances, though concluding whether R2 PanSTARRS is dust-rich or gas rich depends on which volatile is used as the reference.  The value of $log$[$Af\rho$/Q(H$_2$O)] determined for R2 PanSTARRS is larger by a factor of 5-10 than most comets observed at R$_h$ $\sim$ 3 AU in the survey by~\cite{AHearn1995}, meaning that when H$_2$O is the comparison gas R2 PanSTARRS is considered quite dusty.  However, $log$[$Af\rho$/Q(CO$_2$)] is at the low end of comets observed with \textit{Spitzer} and NEOWISE~\citep[][Kelley et al. in prep]{Bauer2015}, and approximately a factor of 10 lower than any Oort Cloud comet in those samples.  So compared to CO$_2$ R2 PanSTARRS is considered a gas-rich comet.  While no systematic study comparing Af$\rho$ to CO has been done, the extremely high CO production rate suggests that if CO is the reference gas, R2 PanSTARRS is incredibly gas-rich, possibly the most gas-rich comet ever observed.  Even in terms of the total gas production (typically dominated by H$_2$O but dominated by CO in the case of R2 PanSTARRS) compared to Af$\rho$, R2 PanSTARRRS is very gas-rich. 

\begin{table}[h!]
\begin{center}
\caption{\textbf{Additional Mixing Ratios for R2 PanSTARRS}
\label{otherabundances}
}
\begin{tabular}{ccccccccccccc}
\hline
Species & X/CO$_2$ (\%) & X/CH$_4$ (\%) &  X/CH$_3$OH (\%) &  X/H$_2$CO (\%) &  X/HCN (\%) \\
\hline
CH$_4$ & 5.8 $\pm$ 0.9 (19) & - & & &\\
CH$_3$OH & 11.6 $\pm$ 1.4 (13) & 200 $\pm$ 28 (250) & - & &\\
H$_2$CO & 0.47 $\pm$ 0.09 (1.9) & 8.0 $\pm$ 1.6 (38) & 4.0 $\pm$ 0.7 (15) & - & \\
HCN & 0.04 $\pm$ 0.01 (1.3) & 0.71 $\pm$ 0.20 (25) & 0.36 $\pm$ 0.09 (10) & 8.9 $\pm$ 2.6 (67) & - \\
C$_2$H$_6$ & $<$ 0.89 (3.7) & $<$ 15 (72) & $<$ 7.6 (29) & $<$ 190 (190) & $<$ 2125 (290)\\
C$_2$H$_2$ & $<$ 0.21 (0.94) & $<$ 3.8 (18) & $<$ 1.9 (7.2) & $<$ 47 (48) & $<$ 525 (72)\\
NH$_3$ & $<$ 0.1 (5.4) & $<$ 1.8 (103) & $<$ 0.9 (41) & $<$ 22 (276) & $<$ 250 (414)\\
\end{tabular}
\end{center}
\end{table}

\subsubsection{Comets at Large Heliocentric Distance}
\indent A caveat to consider when interpreting observations of R2 PanSTARRS is the fairly large heliocentric distance of the observations, $\sim$ 2.8 AU.  We compare to other comets observed at large heliocentric distance (approaching or beyond 3 AU) in Table~\ref{largeR}.  While these comets do show depletions compared to CO, R2 PanSTARRS shows heavier depletions by at least a factor of two for all species reported in Table~\ref{largeR}.  No comet in Table~\ref{largeR} shows an H$_2$O/CO ratio $<$ 20\%, a limit two orders of magnitude higher than the observed value in R2 PanSTARRS, and most have H$_2$O/CO $\gtrsim$ 100\%.  It should also be noted that there is a tendency for comets observed at large heliocentric distance to be highly active, which is potentially attributed to being CO-rich.  Therefore the general overabundance of CO compared to other volatiles for comets observed at large heliocentric distance could simply be due to this observational bias.  29P/Schwassman-Wachmann 1 shows the most similarities with R2 PanSTARRS, with an H$_2$O/CO ratio $\sim$ 20\%, low abundances of CH$_4$ and C$_2$H$_6$ compared to CO, and a reported N$_2$/CO ratio of $\sim$ 1\%~\citep{Ivanova2016}.  However, at 6.2 AU the CO sublimation rate is expected to be $\sim$ 10,000 times faster than the H$_2$O sublimation rate~\citep{CowanAHearn1979}, so accounting for this sublimation effect implies that the intrinsic H$_2$O/CO ratio in the nucleus of 29P may be more in line with that typically observed in comets (H$_2$O/CO $>$ 300\%) \citep{WomackDistantReview2017}.  The upper limits on species such as CH$_4$/CO and HCN/CO are not sensitive enough to show whether 29P is similar to R2 PanSTARRS, as these upper limits are at least one order of magnitude larger than the observed abundances in R2 PanSTARRS.\\
\indent Probably the best studied comet at large heliocentric distance is 67P/Churyumov-Gerasimenko, for which \textit{in situ} measurements with the ROSINA mass spectrometer aboard Rosetta are available~\citep{LeRoy2015,Rubin2015b}.  Values for the winter and summer hemispheres at 3.14 AU pre-perihelion are presented in Table~\ref{largeR}.  Similar to the other comets in Table~\ref{largeR}, both hemispheres exhibit abundances different from R2 PanSTARRS by at least a factor of three, with H$_2$O, CO$_2$, HCN, and C$_2$H$_6$ exhibiting the most striking differences.  The N$_2$/CO ratio for 67P is approximately a factor of 10 smaller than the value for R2 PanSTARRS~\citep{Rubin2015b}, which combined with the CO/H$_2$O ratios given in Table~\ref{largeR} implies an N$_2$/H$_2$O ratio for R2 PanSTARRS four to five orders of magnitude larger than that for 67P.\\   
\indent Moderate enhancements in CO compared to H$_2$O are expected based on the sublimation model of~\cite{CowanAHearn1979}, which predicts that the different volatilities of CO and H$_2$O can account for a factor of two at a heliocentric distance of 2.8 AU, similar to the enhancement observed in other comets near 3 AU but not the factor of $\sim$ 10,000 observed for R2 PanSTARRS.  Therefore the anomalous mixing ratios cannot be ascribed to the heliocentric distance alone and must reflect the intrinsic composition of the nucleus.
\begin{table}[h!]
\begin{center}
\caption{\textbf{R2 PanSTARRS and Other Comets Observed at Large Heliocentric Distance}
\label{largeR}
}
\begin{tabular}{lcccccccc}
\hline
Comet & R$_h$ & Q$_{CO}$ & \multicolumn{6}{c}{Mixing Ratios X/CO (\%)}\\
\cline{4-9}
& (AU) & (10$^{27}$ mol s$^{-1}$) & H$_2$O & CO$_2$ & CH$_4$ & C$_2$H$_6$ & HCN & CH$_3$OH\\
\hline
C/2006 W3$^a$ & 3.13 & 19.8 $\pm$ 2.0 & 102 $\pm$ 15 & 43 $\pm$ 6 & - & - & - & -\\
C/2006 W3$^b$ & 3.26 & 55.5 $\pm$ 2.6 & - & - & 4.4 $\pm$ 0.4 & 2.1 $\pm$ 0.2 & - & -\\
C/2006 W3$^c$ & 3.20 & 39.0 $\pm$ 3.0 & - & - & - & - & 0.42 $\pm$ 0.04 & 3.8 $\pm$ 0.8\\
29P$^a$ & 6.18 & 29.1 $\pm$ 2.1 & 21.6 $\pm$ 2.4 & $<$ 1.1 & - & - & - & -\\
29P$^d$ & 6.26 & 26.7 $\pm$ 2.8 & $<$ 1830 & - & $<$ 4.8 & $<$ 2.2 & $<$ 5.2 & $<$ 35.3\\
C/2010 G2$^e$ & 2.50 & 4.7 $\pm$ 0.7 & 93 $\pm$ 20 & - & 9.0 $\pm$ 0.9 & 3.8 $\pm$ 0.4 & - & -\\
C/1995 O1$^f$ & 2.93 & 230 & 140 & 32 & - & - & - & -\\
C/1995 O1$^g$ & 2.84 & 200 & 300 $\pm$ 39 & - & - & - & 0.58 $\pm$ 0.10 & 7.1 $\pm$ 0.7\\
C/2006 OF2$^a$ & 3.2 & $<$ 0.45 & $>$ 377 & $>$ 220 & - & - & - & -\\
C/2006 Q1$^a$ & 2.78 & $<$ 0.37 & $>$ 973 & $>$ 432 & - & - & - & - \\
C/2007 G1$^a$ & 2.80 & $<$ 0.32 & $>$ 563 & $>$ 131 & - & - & - & -\\
C/2007 Q3$^a$ & 3.29 & $<$ 0.40& $>$ 1000 & $>$ 175& - & - & - & -\\
C/2008 Q3$^a$ & 2.96 & $<$ 0.39 & $>$179 & $>$ 115 & - & - & - & -\\
67P$^h$ & 3.14 & $\sim$ 0.01$^i$ & 3700 & 93 & 4.8 & 11.9 & 3.3 & 11.5\\
67P$^j$ & 3.14 & $\sim$ 0.01$^i$ & 500 & 400 & 2.8 & 16.5 & 3.1 & 2.8\\
\hline
C/2016 R2$^k$ & 2.8 & 95.4 $\pm$ 9.1 & 0.32 $\pm$ 0.04 & 18.2 $\pm$ 3.5 & 0.59 $\pm$ 0.09 & $<$ 0.089 & (3.8 $\pm$ 1.0) $\times$ 10$^{-3}$ & 1.04$\pm$ 0.08\\
\end{tabular}
\end{center}
$a$ \cite{Ootsubo2012}\\
$b$ \cite{Bonev2017}\\
$c$ \cite{BockeleeMorvan2010}\\
$d$ \cite{Paganini2013}\\
$e$ \cite{Kawakita2014}, spectra obtained after outburst\\
$f$ \cite{Crovisier1997b}\\
$g$ \cite{Biver2018}\\
$h$ \cite{LeRoy2015}, Summer Hemisphere\\
$i$ \cite{LeRoy2015} do not provide production rates, therefore we adopt the approximate CO production rate from~\citep{Fougere2016}, their Fig. 11.\\
$j$ \cite{LeRoy2015}, Winter Hemisphere\\
$k$ Values are those shown in Table~\ref{abundances}\\
\end{table}

\subsection{Active Fractions}
\indent Our derived active areas and active fractions for H$_2$O, CO$_2$, and CO are given in Table~\ref{activearea}.  While the nucleus size of R2 PanSTARRS is not known, the very large active area required for CO would require the nucleus to be quite large ($>$ 5 km) for all the CO to come from surface sublimation.  There is also the possibility of an extended source of CO, which adds additional available surface area for CO to sublimate from.  However, our \textit{Spitzer} data (which is sensitive to both CO and CO$_2$ as discussed earlier) do not show evidence for increasing production rate with photometric aperture size, as would be expected for an extended source~\citep{Combi2013, Bodewits2014, McKay2015}, arguing against an extended source larger than $\sim$ 10,000 km in radius (corresponding to $\sim$ 5 pixels in our Spitzer images, the smallest photometric aperture for which reliable photometry can be performed) for these molecules.  We cannot definitively rule out the possibility of an extended source of smaller spatial extent than our Spitzer photometric apertures, though our modeling of the CO spatial profile in our iSHELL observations using only optical depth effects (see Section~\ref{subsec:ishellda}) provides some evidence against a smaller extended source.\\
\indent The derived H$_2$O active fraction is consistent with other comets if the nucleus is fairly small ($<$ 3 km), but this would contradict the large active area needed to explain the CO production.  For a large nucleus ($>$ 10 km) this would imply a very low active fraction for H$_2$O sublimation, much lower than other comets~\citep{SosaFernandez2011,Lis2019}.  This is additional evidence that the low water production is not simply an artifact of the large heliocentric distance of R2 PanSTARRS at the time of observation, but is part of the inherent composition of this comet.    
\subsection{Implications}
\indent R2 PanSTARRS has an extremely anomalous composition compared to other comets observed to date.  In the previous sections we have demonstrated that the large heliocentric distance can only explain a small portion of the observed composition.  Additionally, the depletion of highly volatile species like CO$_2$ and CH$_4$ compared to CO cannot be explained by the heliocentric distance either.  Therefore the observed anomalous abundances are not a consequence solely of the heliocentric distance.  It is not likely due to thermal evolution from repeated solar passages, as this would work to deplete the most volatile species like CO, N$_2$, CH$_4$, and CO$_2$, not enhance them as observed.  Therefore the composition of R2 PanSTARRS likely reflects its composition when it was formed.\\
\indent CO and N$_2$ are the most primitive molecular forms of carbon and nitrogen, respectively, in the universe.  They often are the starting point of chemical pathways that result in the formation of more complex molecules such as CH$_3$OH, NH$_3$ and HCN.  Therefore the large abundance of CO and N$_2$ in R2 PanSTARRS compared to these more complex molecules suggests that the region of the disk where R2 PanSTARRS formed was chemically inactive, and shielded from photodestruction, leaving the volatile carbon and nitrogen in their simplest forms.  Both CO and N$_2$ are also extremely volatile, as are CH$_4$ and CO$_2$.  The presence of these molecules suggests that R2 PanSTARRS must have formed in the farthest reaches of the protosolar disk in order to retain these hypervolatiles.  The presence of these hypervolatiles also suggests little depletion of these ices from repeated solar passages, despite R2 PanSTARRS likely not being dynamically new and therefore having likely experienced at least several passages through the planetary region (A dynamical analysis by \cite{Opitom2019} found that 100\% of their 1000 R2 clones experienced at least three perihelion passages of less than 3 AU).~\cite{WierzchosWomack2018} propose that the CO, N$_2$ and HCN relative abundances in the coma may be explained by the comet forming in an environment $\sim$ 50K (though other models suggest N$_2$ requires colder temperatures around 20K to condense out of the gas phase~\citep{Drozdovskaya2016}) with significant shielding for N$_2$. Our observations showing very high N$_2$, with very low NH$_3$ abundances may provide additional support for this model, and thus the comet may more closely resemble the composition of the nitrogen-bearing volatiles of dense molecular clouds \citep{Womack1992N2NH3} and YSO's~\citep{Gibb2004}.  The large abundance of CO and CO$_2$ in R2 PanSTARRS is typical of interstellar apolar ice mantles, and high N$_2$ abundances are also expected in these ice mantles~\citep{Gibb2004}.  However, interstellar ice grains also have an abundant polar component that has a large water ice abundance, resulting in H$_2$O ice being the dominant component of ISM ice grains~\citep{Gibb2004}, unlike what we observe for R2 PanSTARRS.\\
\indent Of the observed species, our measurements suggest that the dominant carbon-bearing molecules are CO and CO$_2$, while the dominant nitrogen-bearing molecule is N$_2$.  While CO and CO$_2$ are the main reservoirs of volatile carbon in comets, NH$_3$ and to a lesser extent HCN are typically the dominant reservoirs of volatile nitrogen in comets.  Assuming CO and CO$_2$ contain the majority of the volatile carbon and N$_2$ the volatile nitrogen in R2 PanSTARRS implies a C/N ratio of $\sim$11, while most comets, for which NH$_3$ is the main volatile nitrogen carrier, have a higher C/N ratio of $\sim$20.  The solar value for C/N is $\sim$ 3.4~\citep{Lodders2010}, so while R2 PanSTARRS has a C/N ratio closer to solar than most comets observed to date, its coma is still deficient in nitrogen compared to the Sun.\\
\indent Our results show that most of the volatile oxygen in R2 PanSTARRS is locked in CO and CO$_2$, not H$_2$O as is typically the case.  This may suggest that R2 PanSTARRS formed in a region of the protosolar nebula where the C/O ratio $>$ 1, as chemical models predict that when carbon is more abundant than oxygen in the gas phase most oxygen will be locked into CO and CO$_2$, leaving little oxygen to form H$_2$O.  However, there is also evidence from both comets and protosolar disk models that most of the water in the Solar System was inherited from the presolar cloud rather than formed in the protosolar disk~\citep{Cleeves2014,Altwegg2017}.  If accurate, then the lack of H$_2$O in R2 PanSTARRS could reveal details of how inherited H$_2$O was distributed throughout the protosolar disk, i.e. whether H$_2$O was distributed heterogeneously throughout the disk.  Another possible reservoir for volatile oxygen is O$_2$, which was detected with a surprisingly large abundance by the Rosetta spacecraft at comet 67P/Churyumov-Gerasimenko~\citep{Bieler2015, Keeney2017} and also in archival data from the Giotto spacecraft at 1P/Halley~\citep{Rubin2015}.  O$_2$ is extremely difficult to detect remotely, and none of our observations are sensitive to O$_2$, so we cannot rule out the presence of a large amount of O$_2$ in R2 PanSTARRS.  Production of O$_2$ at 67P was found to correlate well with H$_2$O production~\citep{Bieler2015, Fougere2016}, and because of this some theories for the origin of O$_2$ in comets invoke a strong tie to H$_2$O, either through radiolysis or trapping in clathrates~\citep[e.g.][]{Mousis2016, Dulieu2017, Laufer2017}.  Given the very low H$_2$O abundance in R2 PanSTARRS, this would suggest the O$_2$ abundance should also be very low.  However, given the very peculiar chemistry of R2 PanSTARRS and our limited understanding of O$_2$ incorporation into cometary nuclei, we do not consider this argument definitive proof against a substantial O$_2$ abundance in R2 PanSTARRS.\\
\indent All these implications only apply to the volatile component of R2 PanSTARRS.  There are no constraints on the composition of the refractory component (i.e. dust), so we cannot make any conclusions about atomic abundances in the bulk (i.e. dust and ice) composition of R2 PanSTARRS.\\
\indent \cite{Biver2018} suggest that R2 PanSTARRS may be a fragment of a differentiated Kuiper Belt body in order to explain the large observed hypervolatile abundances.  The relative abundances of CO, CH$_4$, and N$_2$ we observe for R2 PanSTARRS do not match surface spectra of Pluto~\citep{Protopapa2008}, though the relationship between the surface composition and interior of large KBO's like Pluto is unclear.  A detailed analysis of the dynamics of creating collisional fragments in the Kuiper Belt after differentiation and then dynamically transporting these fragments to the Oort Cloud (as well as the expected volatile composition of such fragments) must be further investigated.\\
\indent Such an anomalous composition brings up the possibility that R2 PanSTARRS has an interstellar origin.  While the current orbit of R2 PanSTARRS does not suggest an interstellar origin, R2 PanSTARRS could be a comet captured from another Oort Cloud in the Sun's birth cluster (or a more distant planetary system, though this is less likely) during the Solar System's earliest stages.  Interstellar origins have also been suggested for other comets with peculiar compositions such as 96P/Machholz~\citep{Schleicher2008} and C/1988 Y1 (Yanaka)~\citep{Fink1992}. It has been shown that there could have been exchange of comets between Oort Clouds in the Sun's birth cluster~\citep{Levison2010}.  However, it is not clear whether these comets would be expected to be significantly different compositionally from ``solar'' comets as both they and their host stars would have formed from the same nebular gas.  
\section{Conclusion} \label{sec:end}
We present IR, optical, and millimeter wavelength observations of comet C/2016 R2 (PanSTARRS) and show it has a very peculiar composition compared to typical comets, with strong enhancements in species such as CO and N$_2$ and strong depletions in species such as H$_2$O and HCN, as revealed by previous studies.  We determined through observations of a suite of 12 species that the anomalous composition of R2 PanSTARRS is not reserved for one or two species, but exhibits strong deviations from typical comets for most species and reference points (i.e comparing to H$_2$O, CO, CH$_4$, CH$_3$OH, etc.). The lone exceptions are CH$_3$OH/CO$_2$ and CH$_3$OH/CH$_4$, which are considered typical.  We also show that the peculiar composition of R2 PanSTARRS is not due solely to the large heliocentric distance at the time of observation, and is intrinsic to the comet.  What implications R2 PanSTARRS has for our knowledge of the early Solar System are still unclear.  We suggest some future lines of research that could shed light on this issue.\\
\\
1) Dynamical modeling of R2 PanSTARRS: This includes both its recent history as investigated by~\cite{Opitom2019}, but also dynamical simulation of the Scattered Disk to evaluate the suggestion of~\cite{Biver2018} that R2 PanSTARRS could be a collisional fragment from a differentiated Kuiper Belt Object.  Models would need to evaluate the likelihood of forming a differentiated KBO, collisionally fragmenting this KBO, then dynamically transporting this fragment to the Oort Cloud.\\
\\
2) Chemical modeling of protosolar disks: We believe that if R2 PanSTARRS is not a collisional fragment of a differentiated KBO, it provides a new constraint on chemical models of protosolar disks.  H$_2$O in the early Solar System was likely inherited from the parent molecular cloud~\citep{Cleeves2014}. Could this create a depletion mechanism for H$_2$O in certain regions of the protosolar disk where R2 PanSTARRS formed?  Is there a region of the protosolar disk where CO and N$_2$ are not efficiently processed into more complex species in order to account for the strong enhancement of these species in R2 PanSTARRS?  These are only a couple questions that chemical modeling of protoplanetary disks and observations of other protoplanetary disks can help answer.\\
\\
3) Continued compositional studies of comets: R2 PanSTARRS is a case-in-point for the importance of remote sensing observations to measure the composition of as many comets as possible.  R2 PanSTARRS was a not particularly bright comet and was at a large heliocentric distance, meaning it could have easily been missed by compositional studies.  More observations of comets are needed to determine how common comets like R2 PanSTARRS are.  As pointed out by~\cite{Biver2018}, the closest comparisons in the historical literature are C/1908 R1 (Morehouse) and C/1961 R1 (Humason), but neither of these were studied with modern capabilities, therefore very little is known about their overall composition.  With current sky surveys like PanSTARRS and LSST coming on line in the coming years, the prospect for discovering more comets like R2 PanSTARRS grows.  The frequency of objects like R2 PanSTARRS will provide meaningful constraints on its history and the formation of our Solar System.

\acknowledgments
We are grateful to the anonymous reviewer for helpful comments that improved the quality of this manuscript.  We thank \textit{Spitzer}, IRTF, Apache Point Observatory, and the Discovery Channel Telescope for granting us DDT or ToO time to conduct the observations described in this paper.  We thank Svetlana Jorstad for graciously allowing us to interrupt her program with our DCT ToO observations.  These results made use of the Discovery Channel Telescope at Lowell Observatory. Lowell is a private, non-profit institution dedicated to astrophysical research and public appreciation of astronomy and operates the DCT in partnership with Boston University, the University of Maryland, the University of Toledo, Northern Arizona University and Yale University. The Large Monolithic Imager was built by Lowell Observatory using funds provided by the National Science Foundation (AST-1005313).  The SMT is operated by the ARO, the Steward Observatory, and the University of Arizona, with support through the NSF University Radio Observatories program grant AST-1140030. This work was completed with the GILDAS CLASA software: \url{http://www.iram.fr/IRAMFR/GILDAS}.  The authors recognize and acknowledge the very significant cultural role and reverence that the summit of Maunakea has always had within the indigenous Hawaiian community. We are most fortunate to have the opportunity to conduct observations from this mountain. AJM acknowledges support from the NASA Postdoctoral Program, administered by the Universities Space Research Association.  MD acknowledges support through NASA grant 15-SSO15$_-$2-0028.  MMK acknowledges support from  NASA Solar System Observations Program grant 80NSSC18K0856.  MW and KW acknowledge support from NSF grant AST-1615917. OHP acknowledges support from the USF Genshaft Family Doctoral Fellowship.  BPB acknowledges support from NSF grant AST-1616306. BPB, NDR, and RJV acknowledge support from NASA Solar System Observations grant 80NSSC17K0705.  NR acknowledges support from the NASA Earth and Space Science Fellowship Program (Grant NNX16AP49H). ALC acknowledges support from the NASA Solar System Observations Program (NNX17A186G).

\vspace{5mm}
\facilities{\textit{Spitzer} IRAC, NASA IRTF iSHELL, ARC 3.5-meter, Discovery Channel Telescope, Arizona Radio Observatory (SMT)}

\bibliography{references.bib}
\bibliographystyle{plainnat}



\end{document}